\documentclass{article}
\usepackage[margin=3cm]{geometry}

\usepackage{upgreek}

\usepackage{authblk}

\usepackage{amsmath,amssymb,amsfonts}
\usepackage[nocompress]{cite}
\usepackage{todonotes}
\numberwithin{equation}{section}

\usepackage{hyperref}
\hypersetup{colorlinks=true,allcolors=blue,urlcolor=cyan}

\title{Consistent energy-momentum trace couplings of fluids}
\author[1,2]{Christian G B\"ohmer\footnote{Email: c.boehmer@ucl.ac.uk}}
\author[1,3]{Eissa Al-Nasrallah\footnote{Email: eissa.alnasrallah.24@ucl.ac.uk}}
\affil[1]{Department of Mathematics, University College London, \authorcr 
Gower Street, London WC1E 6BT, UK\medskip}
\affil[2]{Astrophysics Research Centre, School of Mathematics, \authorcr 
Statistics and Computer Science, University of KwaZulu-Natal, \authorcr 
Private Bag X54001, Durban 4000, South Africa\medskip}
\affil[3]{Department of Physics, College of Science, Kuwait University, \authorcr 
Sabah Al Salem University City, P.O. Box 2544, Safat 1320, Kuwait}

\date{\today} 

\begin{document}
\maketitle

\begin{abstract}
Gravitational models with nonminimal couplings involving the trace of the energy-momentum tensor have become increasingly popular. The idea of coupling the trace of the matter tensor to the geometry can be applied to various matter models, including relativistic perfect fluids. However, it is well known that the variational formulation of perfect fluids involves some technicalities. We carefully derive the field equations including the trace coupling of a perfect fluid using two different approaches, namely, that given by Brown using Lagrange multipliers and that given by Schutz using velocity potentials. We show that previous results involving such trace couplings do not match the results presented here. We demonstrate that our fluid's equations of motion are consistent with the gravitational field equations. Moreover, we present a simple on-shell argument which further supports the correctness of our results. 
    
Our work implies that a vast amount of the $f(R,T)$ literature using perfect fluids needs to be revisited.
\end{abstract}

\clearpage 
\tableofcontents
\clearpage

\section{INTRODUCTION}

Recent advances in cosmology have revealed that new physics beyond general relativity is necessary to explain observed phenomena. There is a plethora of modified theories of gravity which include alternative geometries with torsion and nonmetricity, higher-order theories, and scalar and matter couplings~\cite{CANTATA:2021asi}. Nonminimal matter-curvature couplings are generalizations of $f(R)$ gravity in which an additional matter term is coupled with the Ricci scalar $R$ in the action. The first of these models appeared in the work of Bertolami \textit{et al.}~\cite{Bertolami:2007gv}, where curvature is coupled with Lagrangian density of matter in the form of $f(R,L)$. Since then, these models have been found particularly useful in providing an alternative account of the Universe expansion or dark energy, and dark matter\cite{Lobo:2008sg}. Matter-curvature nonminimal couplings do not conserve the energy-momentum tensor. The covariant derivative of the energy-momentum tensor often results in a nonvanishing term that acts as an additional force term which violates geodesic motion. Thus, these theories have been shown promising in resolving cosmological issues in the weak field regime associated with dark sector physics \cite{harko_extensions_2019}. 

In 2011 a new idea was proposed in~\cite{Harko:2011kv} which suggested to couple the Ricci scalar to the trace of the energy-momentum tensor. This has been known as $f(R,T)$ gravity and became hugely influential with over 2300 citations to date, according to INSPIRE. The trace of the energy-momentum tensor is itself a function of not only the Lagrangian, but also its variations with respect to the metric. The field equations are correctly stated in~\cite{Harko:2011kv} in terms of a function that depends on the variation of matter Lagrangian; see also~\cite{Haghani:2023uad}. However, when applied to a perfect fluid, the situation is rather delicate when using $-\rho\sqrt{-g}$ or $p\sqrt{g}$ as the fluid Lagrangian. Specifically, one has to be careful with the derived quantity, either the density or the pressure, in order to arrive at the correct field equations when nonminimal couplings are considered.

When working with relativistic perfect fluids, one notices that there is no unique definition of the matter Lagrangian. Several approaches have been shown to lead to the expected energy-momentum tensor of the fluid. The energy-momentum tensor of a relativistic perfect fluid is characterized by the unit four-velocity vector of the flow $U^\nu$, the energy density $\rho$, and pressure $p$ and is given by
\begin{align}
      T_{\mu\nu} = (\rho+p) U_\mu U_\nu + p g_{\mu\nu} \,,
\end{align}
where $g_{\mu\nu}$ is the metric tensor; see for example \cite{Landau:1980mil,Misner:1973prb}.

On shell, meaning that all equations of motion have been taken into account, one does indeed find that $\mathcal{L}_{\rm fluid} = -\rho \sqrt{-g} = p \sqrt{-g}$. However, the situation is more complicated when nonminimal coupling is present (see~\cite{Bertolami:2008ab, Avelino:2022eqm}). In a variational approach, one must start with the complete off-shell Lagrangian which will lead to the correct field equations and the fluid's equations of motion. Only then can one use the equations of motion in the Lagrangian to get the on-shell expression. A detailed account on the variational approach to relativistic perfect fluids is given in~\cite{Brown_1993,Schutz:1970}. The consideration of the off-shell Lagrangian for the perfect fluid becomes particularly important when one derives the fluid's equations of motion. In $f(R,T)$ gravity, the variation of the trace of the energy-momentum tensor introduces new contributions that, as will be shown, modify the expressions for the temperature, number density, and chemical potential. This allows us to define the temperature contribution of the nonminimal coupling, for example, an important physical quantity not previously discussed. 

In~\cite{Brown_1993}, Brown uses an action functional that depends on Lagrange multipliers and Lagrange coordinates. It introduces Lagrangian coordinates $\alpha^A$ that define the flow lines of the fluid on a spacelike hypersurface. The fluid's energy density is given as a function of the particle-number density $n$ and entropy $s$, while the pressure $p$ is a derived quantity from the density and its derivative. The Lagrangian is then defined as a function of the aforementioned parameters along with particle-number flux $J^\nu$ and additional spacetime scalars. This formalism provides a Lagrangian that explicitly depends on the independent parameters which, although cumbersome, allows us to account for all the variational effects.

In an earlier paper, Schutz presented the relativistic perfect fluid Lagrangian using a velocity potentials approach in an Eulerian picture~\cite{Schutz:1970}. In this picture, one can identify the Lagrangian as $\mathcal{L}_{\rm fluid}=p\sqrt{-g}$, which looks like the on-shell expression. However, the pressure $p$ here is a function of the chemical potential $\upmu$ and entropy $s$. Contrary to Brown's approach, the density here is a derived quantity from the pressure. The four-velocity vector in this approach is now a function of six potentials and their gradients and thus, in the variational approach, one has to carefully account for the variation of each independent potential. These are $\upmu$ and $s$ along with four additional potentials $\{\phi, \beta,\alpha,\theta\} $. We will present the action functional variation in both Brown's and Schutz's approaches to show the analogy between the results in each and to verify our approach using different Lagrangians.

We start with Brown's Lagrange multipliers picture in Sec.~\ref{sec_Brown}, with the same analysis done in Schutz's velocity potential picture in Sec.~\ref{sec_Schutz}. In each section, we begin by deriving the field equations and equations of motion. We show that the equations lead to the emergence of effective quantities which allow us to derive the on-shell Lagrangian and introduce new thermodynamic quantities. Finally, we derive the conservation equations starting first from the field equations and then from the fluid equation and demonstrate their equivalence. We shall then see in Sec.~\ref{sec_implications} that the results of both pictures lead to the same implications. Specifically, one finds that the nonminimal coupling in $f(R,T)$ merely affects the equation of state when models linear in $T$ are considered.

\section{BROWN'S LAGRANGE MULTIPLIER APPROACH}
\label{sec_Brown}

Brown's approach has the advantage of explicitly stating the independent variables in the Lagrangian density function~\cite{Brown_1993}. One can clearly track the variation of each variable which makes this approach more convenient to use. The Lagrangian density is given by
\begin{align}
\label{eq_lagrangian_fluid}
    \mathcal{L}_{\mathrm{fluid}} = -\rho(n,s)\,\sqrt{-g} + J^\mu(\varphi_{,\mu}+s\theta_{,\mu}+\beta_A\alpha^A_{,\mu}) \,,  
\end{align}
where
\begin{itemize}
    \item $n$ is the particle-number density;
    \item $s$ is entropy density per particle;
    \item $\rho(n,s)$ is the energy density of the matter fluid, a function of $n $ and $s$;
    \item  $J^\mu$ is the vector-density particle-number flux, which is related to $n$ by
    \begin{equation}
        J^\mu=\sqrt{-g}nU^\mu\,, \qquad 
        |J|=\sqrt{-g_{\mu\nu}J^\mu J^\nu}\,, \qquad 
        n=|J|/\sqrt{-g} \,, 
        \label{defnofJandn}
    \end{equation}
    and $U^\mu$ is the fluid's four-velocity satisfying $U_\mu U^\mu=-1$;
    \item  $\varphi$, $\theta$, and $\beta_A$ are all Lagrange multipliers with $A$ taking the values 1, 2, and 3, and $\alpha^A$ are the Lagrangian coordinates of the fluid; and
    \item the pressure $p$ of the fluid is a derived quantity defined by 
    \begin{align}
        p = n\frac{\partial \rho}{\partial n} - \rho \,.
        \label{pressuredef}
    \end{align}
\end{itemize}
 Now, we can obtain the field equations and equations of motion by applying the variational principle on the action, using the Lagrangian density given by Eq.~\eqref{eq_lagrangian_fluid}. The independent variables which we vary with respect to are
 \begin{align}
     \{g^{\mu\nu}, J^\mu, \alpha^A, \beta_A, \varphi, \theta, s\} \,.
 \end{align}
 The variation with respect to $g^{\mu\nu}$ yields the gravitational field equations, while the variations with respect to the other variables yield the fluid's equations of motion.
 
\subsection{Gravitational field equations}
\label{Brown Gravitational field equations}

We begin by carefully deriving the full variation of the action first and then derive the field equations from the full variation. The equations of motion of the fluid, also obtained from the full action variation, are derived in the next section.

As in~\cite{Harko:2011kv} we begin with the $f(R,T)$ modified gravity model
\begin{align}
\label{eq_action}
    S = \frac{1}{2\kappa}\int f(R,T)\sqrt{-g}\,d^4x + 
    \int \mathcal{L}_{\rm fluid} d^4x \,.
\end{align}
The variation of the action is
\begin{align}
    \label{eq_S_variation}
    \delta S &= \frac{1}{2\kappa}\int \Bigl[
    \frac{\partial f}{\partial R}\delta R \sqrt{-g} +
    \frac{\partial f}{\partial T}\delta T \sqrt{-g} - 
    f\delta\sqrt{-g}\Bigl]  d^4x + 
    \int \delta\mathcal{L}_{\rm fluid}d^4x \,.
\end{align}
where $f$ is understood to be $f(R,T)$. From this point onward, we also write $f_R = \partial f/\partial R$ and likewise $f_T = \partial f/\partial T$. The variation of the square root of the determinate of the metric is $\delta \sqrt{-g} = -\frac{1}{2}\sqrt{-g}g_{\mu\nu}\delta g^{\mu\nu} $. The expression of $\delta R$ is well known and important in the derivation of the field equations of $f(R)$ gravity. It suffices to state the result directly:
\begin{align}
    \label{eq_R_variation}
    \delta R = \big(R_{\mu\nu} + g_{\mu\nu}\square - \nabla_\mu \nabla_\nu \big)\delta g^{\mu\nu} \,.
\end{align}
The variations $\delta T$ and $\delta \mathcal{L}_{\rm fluid}$ are the matter variations which are now done in the context of Brown's fluid. 

\subsubsection{Variation of $T$}

As this calculation has not yet been performed in Brown's fluid approach, we provide a detailed derivation here. We start from the well-known equation for the energy-momentum tensor for a relativistic perfect fluid:
\begin{align}
    T_{\mu\nu} = (\rho+p) U_\mu U_\nu + p g_{\mu\nu} \,.
\end{align}
The pressure $p$, as defined earlier, is a derived quantity and not independent. This allows us to use Eq.~\eqref{pressuredef} to write the trace as
\begin{align}
\label{eq_trace_em}
    T = g^{\mu\nu}T_{\mu\nu} = (\rho+p)U^\mu U_\mu + 4 p = 3p - \rho = 
    3\Bigl(n\frac{\partial \rho}{\partial n} - \rho \Bigr) - \rho =
    3 n\frac{\partial \rho}{\partial n} - 4\rho \,.
\end{align}
We now vary the trace of the energy-momentum tensor and recall that the energy density is a function of the particle-number density and entropy $\rho=\rho(n,s)$:
\begin{align}
    \delta T &= 3 \frac{\partial \rho}{\partial n} \delta n + 
    3 n\frac{\partial^2 \rho}{\partial n^2} \delta n +
    3 n\frac{\partial^2 \rho}{\partial n \partial s} \delta s
    - 4\frac{\partial \rho}{\partial n} \delta n
    - 4\frac{\partial \rho}{\partial s} \delta s \notag \\
    &=
    \Bigl[3 n\frac{\partial^2 \rho}{\partial n^2}
    - \frac{\partial \rho}{\partial n}\Bigr] \delta n +
    \Bigl[3 n\frac{\partial^2 \rho}{\partial n \partial s} - 4\frac{\partial \rho}{\partial s} \Bigr] \delta s \,.
\end{align}
We note that the expressions in the square brackets are the partial derivatives of the trace with respect to $s$ and $n$:
\begin{align}
    \frac{\partial}{\partial s}T &= 3 n\frac{\partial^2 \rho}{\partial n \partial s} - 4\frac{\partial \rho}{\partial s}\,, \\
    \frac{\partial}{\partial n}T & = 3 n\frac{\partial^2 \rho}{\partial n^2 }  - \frac{\partial \rho}{\partial n} \,,
    \label{usedTdn}
\end{align}
respectively. Consequently, we arrive at the expected result 
\begin{align}
    \delta T &= 
    \frac{\partial T}{\partial n} \delta n +
    \frac{\partial T}{\partial s} \delta s \,.
\end{align}
Notice that while the entropy $s$ is an independent variable, the particle-number density $n$ is not. The variation $\delta n$ can be further expanded in terms of the independent variables (see detailed derivation in Appendix~\ref{appen_var_n}):
\begin{align}
    \delta n =\frac{1}{2}n\Bigl[g_{\mu\nu} +
    U_\mu U_\nu \Bigr] \delta g^{\mu\nu}-
    \frac{1}{\sqrt{-g}} U_\mu\delta J^\mu \,.
\end{align}
This yields the expression for the variation of the trace of the energy-momentum tensor in terms of the independent variables as
\begin{align}
\label{eq_T_var}
    \delta T = 
    \frac{1}{2}n  \frac{\partial T}{\partial n} 
    \Bigl(g_{\mu\nu} +
    U_\mu U_\nu \Bigr) \delta g^{\mu\nu} -
     \frac{\partial T}{\partial n}
    \frac{1}{\sqrt{-g}} U_\mu \delta J^\mu  +
     \frac{\partial T}{\partial s} \delta s \,.
\end{align}
We see that the variation of the trace of energy-momentum tensor yields not only variations with respect to the metric $\delta g^{\mu\nu}$, but also variations of the particle-number flux  $\delta J^\mu$ and  entropy $\delta s$. The appearance of entropy contribution in the action variation is particularly interesting as this does not appear before in~\cite{Harko:2011kv}.

\subsubsection{Variation of $\mathcal{L}_{\rm fluid}$}

The first step in varying the terms of the Lagrangian density given in Eq.~\eqref{eq_lagrangian_fluid} reveals that
\begin{align}
\label{eq_Lagrangian_var_1}
    \delta \mathcal{L}_{\mathrm{fluid}} = -\rho\delta \sqrt{-g} - \sqrt{-g}\delta \rho + (\varphi_{,\mu}+s\theta_{,\mu}+\beta_A\alpha^A_{,\mu})\delta J^\mu + J^\mu \delta(\varphi_{,\mu}+s\theta_{,\mu}+\beta_A\alpha^A_{,\mu}) \,.
\end{align}
We have already derived $\delta\sqrt{-g}$ and $\delta\rho(n,s)$ in terms of the independent variables. We also notice that the last term is written in terms of independent variables and their derivatives only. By varying each term, integrating by parts, and using the definition of $J^\mu$ from Eq.~\eqref{defnofJandn}, we find the expression for the variation of the energy-momentum Lagrangian density: 
\begin{multline}
\label{eq_L_variation}
    \delta \mathcal{L}_{\rm fluid} =
    \frac{1}{2}\sqrt{-g}\Big[g_{\mu\nu}\rho-\frac{\partial \rho}{\partial n}n
    (g_{\mu\nu} + U_\mu U_\nu)\Big]\delta g^{\mu\nu} +
    \Big[\frac{\partial \rho}{\partial n}U_\mu + 
    \varphi_{,\mu}+s\theta_{,\mu}+\beta_A\alpha^A_{,\mu}\Big]\delta J^\mu \\+ \sqrt{-g}\Big[n U^\mu \theta_{,\mu} -
    \frac{\partial \rho}{\partial s}\Big] \delta s  -
    \big(\partial_\mu J^\mu\big) \delta \varphi - \big(\partial_\mu( J^\mu s)\big)\delta\theta + \big(J^\mu\alpha^A_{,\mu}\big)\delta\beta_A - \big(\partial_\mu(J^\mu\beta_A)\big)\delta\alpha^A \,.
\end{multline}
The detailed derivation of the expression is given in Appendixes~\ref{appen_var_n} and~\ref{appen_var_L}. We now have contributions from the Lagrange multipliers variations, which we will later use to derive the equations of motion.

\subsubsection{The full action variation}

We are now ready to substitute the terms in Eqs.~\eqref{eq_R_variation},~\eqref{eq_T_var}, and~\eqref{eq_L_variation} into the action variation in Eq.~\eqref{eq_S_variation}. This leads to the full action variation which contains all the information for $f(R,T)$ modified gravity model, with a relativistic perfect fluid expressed in Brown's picture:
\begin{align}
    \delta S = \frac{1}{2\kappa}\int &\Bigl[
    f_R R_{\mu\nu}
    +g_{\mu\nu}\Box f_R - \nabla_\mu\nabla_\nu f_R - \frac{1}{2}fg_{\mu\nu} \Bigr]\delta g^{\mu\nu} \sqrt{-g}\,d^4x \notag \\
    + \frac{1}{2\kappa}\int &f_T \Big[\frac{1}{2}n\frac{\partial T}{\partial n} 
    (g_{\mu\nu} + U_\mu U_\nu) \delta g^{\mu\nu} -
    \frac{\partial T}{\partial n} 
    \frac{J_\mu}{\sqrt{-g}|J|} \delta J^\mu +
    \frac{\partial T}{\partial s} \delta s\Bigr] \sqrt{-g}\,d^4x \notag \\
    + \int &\biggl[\frac{1}{2}\sqrt{-g}\Bigl[g_{\mu\nu}\rho-\frac{\partial \rho}{\partial n}n
    (g_{\mu\nu} + U_\mu U_\nu)\Bigr]\delta g^{\mu\nu} \notag \\ 
    &+ \Bigl(\frac{\partial \rho}{\partial n}U_\mu + 
    \varphi_{,\mu}+s\theta_{,\mu}+\beta_A\alpha^A_{,\mu}\Bigr)\delta J^\mu + \sqrt{-g}\Bigl(n U^\mu \theta_{,\mu} -
    \frac{\partial \rho}{\partial s}\Bigr)\delta s  \notag \\
    &-(\partial_\mu J^\mu) \delta \varphi - (\partial_\mu( J^\mu s))\delta\theta + (J^\mu\alpha^A_{,\mu})\delta\beta_A - (\partial_\mu(J^\mu\beta_A))\delta\alpha^A\biggr] d^4x \,.
\end{align}
 By setting $\delta S = 0$ and recognizing that the equations should hold for arbitrary variations, we can extract the equations for the variation of each independent variable. First, we collect the $\delta g^{\mu\nu}$ terms to get the field equations
\begin{multline}
    \frac{1}{2\kappa}\Bigl[
    f_R R_{\mu\nu}
    +g_{\mu\nu}\Box f_R - \nabla_\mu\nabla_\nu f_R - \frac{1}{2}fg_{\mu\nu} \Bigr] + \frac{1}{4\kappa}f_T n\frac{\partial T}{\partial n} 
    (g_{\mu\nu} + U_\mu U_\nu) \\ + 
    \frac{1}{2}\Bigl(g_{\mu\nu}\rho-\frac{\partial \rho}{\partial n}n
    (g_{\mu\nu} + U_\mu U_\nu)\Bigr) = 0 \,,
\end{multline}
which can be written in a more familiar form:
\begin{align}
    \label{eq_our_field_eqn}
    f_R R_{\mu\nu} + \big(g_{\mu\nu}\Box - \nabla_\mu\nabla_\nu\big) f_R - 
    \frac{1}{2} f g_{\mu\nu} = 
    \kappa T_{\mu\nu} - \frac{1}{2}\ f_T n\frac{\partial T}{\partial n} 
    (g_{\mu\nu} + U_\mu U_\nu) \,.
\end{align}
Notice here that unlike general relativity and $f(R)$ gravity, the left-hand side of the equation depends on the fluid matter through the dependence on the trace of the energy-momentum tensor. The right-hand side has the familiar $\kappa T_{\mu\nu}$ term along with an extra term that depends on the derivative of $T$ with respect to the particle-number density. One is tempted to define 
\begin{align}
T^{\rm eff}_{\mu\nu} =T_{\mu\nu} - \frac{1}{2\kappa} f_T n\frac{\partial T}{\partial n}(g_{\mu\nu} + U_\mu U_\nu) \,,
\end{align}
which allows us to express the field equations as
\begin{align}
\label{eq_field_eqns_eff}
     f_R R_{\mu\nu} + \big(g_{\mu\nu}\Box - \nabla_\mu\nabla_\nu\big) f_R - 
    \frac{1}{2} f g_{\mu\nu} = \kappa T_{\mu\nu}^{\rm eff} \,.
\end{align}
In fact, a closer look at the $T^{\rm eff}_{\mu\nu}$ reveals that we can also write
\begin{align}
    T^{\rm eff}_{\mu\nu} &=
   (\rho+p)U_\mu U_\nu+pg_{\mu\nu}-\frac{1}{2\kappa} f_T n\frac{\partial T}{\partial n} 
    (g_{\mu\nu} + U_\mu U_\nu) \notag\\
    &= (\rho + p_{\rm eff})U_\mu U_\nu + p_{\rm eff} g_{\mu\nu}  \,.
    \label{TeffBrown}
\end{align}
with the effective pressure defined as
\begin{align}
    \label{eq_p_eff}
    p_{\rm eff} = p - \frac{1}{2\kappa}f_T n\frac{\partial T}{\partial n} \,.
\end{align}
It now appears that the entire effect of the trace coupling has been absorbed into the effective energy-momentum tensor. While this is formally correct, at this point this is misleading as $p_{\rm eff}$ and $T_{\mu\nu}^{\rm eff}$ depend on $f_T$ and thus also on the Ricci scalar unless $f_{TR}=0$. This is an expected result for models with nonminimal coupling as one can no longer separate the geometry from the matter on each side of the equation. We will discuss this result further but let us first present the equations of motion derived from the action variation. 
\subsection{Equations of motion}
We have six independent variables, besides the metric, which means we get six equations of motion. Four of them can easily be read off and match those presented in~\cite{Brown_1993}. We find
\begin{alignat}{2}
    \delta\alpha^A:& \qquad &\partial_\mu(J^\mu\beta_A) &= 0 \,,
    \label{alphaeq} \\
    \delta\beta_A:& \qquad &J^\mu\partial_\mu\alpha^A &= 0 \,,
    \label{betaeq} \\
    \delta\theta:& \qquad &\partial_\mu( J^\mu s) &= 0 \,,
    \label{thetaeq} \\
    \delta\varphi:& \qquad &\partial_\mu J^\mu &= 0 \,.
    \label{phieq}
\end{alignat}
This is an expected result as we expect the conservation of the particle-number density to hold [Eq.~\eqref{phieq}]. The second condition in Eq.~\eqref{betaeq} restricts the four-velocity direction to constant Lagrange coordinates flow lines. From Eqs.~\eqref{alphaeq} and~\eqref{thetaeq} we have restriction on the exchange of $\beta_A$ and $s$ across the flow.

Furthermore, we can see that the variations with respect to entropy and the four-vector field lead to
\begin{alignat}{2}
    \delta s:& \qquad &n U^\mu \theta_{,\mu} -
    \frac{\partial \rho}{\partial s}+\frac{1}{2\kappa}f_T\frac{\partial T}{\partial s} &= 0 \,,
    \label{seq}\\
    \delta J^\mu:& \qquad &\frac{\partial \rho}{\partial n}U_\mu - 
    \frac{1}{2\kappa}f_T \frac{\partial T}{\partial n} U_\mu + 
    \varphi_{,\mu}+s\theta_{,\mu}+\beta_A\alpha^A_{,\mu} &= 0 \,,
    \label{Jeq}
\end{alignat}
respectively. Now, we get interesting results as both of these equations have terms that emerge from the nonminimal coupling with the trace of the energy-momentum tensor. The effects emerging from these modified results will be discussed further in the Sec.~\ref{section_Brown_thermodynamic quantities}.

Equations~\eqref{eq_our_field_eqn} and~\eqref{alphaeq}-\eqref{Jeq} are the complete set of equations derived from the action. In the following we will interpret the various equations focussing on the effects of the nonminimal coupling. In particular, we will provide a variety of arguments to substantiate our claim that Eq.~\eqref{eq_our_field_eqn} is the correct field equation of $f(R,T)$ gravity when $T$ is the trace of a perfect fluid.

\subsection{On-shell fluid Lagrangian}
\label{sec:onshell}

In the introduction, we claimed that expressing the Lagrangian density as $\mathcal{L}_{\rm fluid}=p\sqrt{-g}$ is only correct on shell, where the equations of motion have been taken into consideration. One then asks whether we get back the on-shell expression when the equations of motion are incorporated into the Lagrangian. Let us begin recalling with the fluid Lagrangian, Eq.~\eqref{eq_lagrangian_fluid}, and integrating by parts all the derivative terms. This gives
\begin{align}
    \mathcal{L}_{\mathrm{fluid}} &= -\sqrt{-g}\rho(n,s) - \partial_\mu J^\mu \varphi - \partial_\mu(s J^\mu)\theta +
    \beta_A J^\mu \alpha^A_{,\mu} + \partial_\mu\big(J^\mu(\varphi+s\theta)\big) \,,
\end{align}
where we note that the final term is a total derivative. By using the equations of motion~\eqref{alphaeq}-\eqref{phieq}, we find that on shell, the Lagrangian density reduces to
\begin{align}
    \mathcal{L}_{\mathrm{fluid}} &= -\rho \sqrt{-g} \qquad \text{on shell} \,.
\end{align}
This is a well known result and unaffected by the coupling with the energy-momentum scalar; see~\cite{Brown_1993}.

Instead, let us start from the equation of motion~\eqref{Jeq} and contract it with $U^\mu$. This gives
\begin{align}
    U^\mu\Big(\frac{\partial \rho}{\partial n}U_\mu + 
    \varphi_{,\mu}+s\theta_{,\mu}+\beta_A\alpha^A_{,\mu}\Big)-U^\mu\frac{f_T}{2\kappa}\frac{\partial T}{\partial n} 
    U_\mu &= 0 \,, \nonumber \\
    -\frac{\partial \rho}{\partial n} + U^\mu \big( 
    \varphi_{,\mu}+s\theta_{,\mu}+\beta_A\alpha^A_{,\mu}\big) + \frac{f_T}{2\kappa}\frac{\partial T}{\partial n} 
     &= 0 \nonumber \,, \\ \Rightarrow \qquad
     n\frac{\partial \rho}{\partial n}-\frac{f_T}{2\kappa}n\frac{\partial T}{\partial n}
     &= \frac{J^\mu}{\sqrt{-g}} \big( 
    \varphi_{,\mu}+s\theta_{,\mu}+\beta_A\alpha^A_{,\mu}\big) \,.
\end{align}
Substituting $n \partial \rho/\partial n=\rho+p$ and multiplying both sides by $\sqrt{-g}$ gives
\begin{align}
    \rho\, \sqrt{-g}+p\, \sqrt{-g}-\frac{f_T}{2\kappa}n\frac{\partial T}{\partial n}\, \sqrt{-g}
     &= J^\mu\big( 
    \varphi_{,\mu}+s\theta_{,\mu}+\beta_A\alpha^A_{,\mu}\big) \,,
    \nonumber \\
    \Big(p-\frac{f_T}{2\kappa}n\frac{\partial T}{\partial n}\Big)\, \sqrt{-g}
     &=-\rho\, \sqrt{-g}+ J^\mu\big( 
    \varphi_{,\mu}+s\theta_{,\mu}+\beta_A\alpha^A_{,\mu}\big) \,.
\end{align}  
The right-hand side is nothing but the expression of the fluid Lagrangian in Brown's picture. If we recall that the expression that appears on the left-hand side is the effective pressure $p_{\rm eff}$, defined in~\eqref{eq_p_eff}, consequently, this statement is saying that 
\begin{align}
    \mathcal{L}_{\rm fluid} = p_{\rm eff}\, \sqrt{-g}  \,.
\end{align}
This on-shell result verifies that the effective pressure identified above is in fact expected. The contributions due to trace coupling are all included in the expression for the effective pressure and there should be no other matter contributions due to the trace coupling. 

On shell, we can also equate the two on-shell expressions and move the extra term of the effective pressure into the density, in which case one finds
\begin{align}
    p\, \sqrt{-g} = 
    -\Bigl(\rho-\frac{f_T}{2\kappa}n\frac{\partial T}{\partial n}\Bigr)\, \sqrt{-g} =: -\rho_{\rm eff} \sqrt{-g} \qquad \text{on shell} \,.
    \label{onshellfinal}
\end{align}
This rewriting seems to suggest that there exists a formulation where the energy density of the fluid can be rewritten to absorb the effects of the trace coupling. This is precisely what will happen following Schutz's velocity potentials approach as will be seen in Eq.~\eqref{rhoeffSchutz} in Sec.~\ref{sec_grav_field_eqns_schutz}.

\subsection{Thermodynamic quantities}
\label{section_Brown_thermodynamic quantities}
We now revisit the equations of motion~\eqref{seq} and~\eqref{Jeq} using a thermodynamic approach. The first law of thermodynamics can be expressed locally by
\begin{align}
    d\rho = \upmu dn + n\mathcal{T} ds \,,
\end{align}
where $\upmu=\partial\rho/\partial n =n^{-1}(\rho+p)$ is the chemical potential and $\mathcal{T}=n^{-1}\partial\rho/\partial s$ is the temperature (not to be confused with the symbols with $\mu$ and $T$). If we now substitute the expression for the temperature in Eq.~\eqref{seq}, we can write
\begin{align}
    U^\mu \theta_{,\mu} = \mathcal{T}-\frac{1}{n}\frac{f_T}{2\kappa}\frac{\partial T}{\partial s}=
    \mathcal{T} - \mathcal{T}_{\rm tr} \,.
\end{align}
We see that the effect of the nonminimal trace coupling is a temperature contribution given by
\begin{align}
\label{eq_temp_brown}
     \mathcal{T} _{\rm tr} := \frac{1}{n}\frac{f_T}{2\kappa}\frac{\partial T}{\partial s} \,.
\end{align}
Since $f_T$ is a function of the matter trace and the Ricci scalar, the temperature contribution to the fluid depends also on a geometric contribution. We also note immediately that, for matter with constant entropy, there will be no temperature contribution to the fluid which is expected as constant entropy processes do not allow for heat transfers.  

Similarly, we can introduce the chemical potential in Eq.~\eqref{Jeq} and write it as
\begin{align}
    (\upmu - \upmu_{\rm tr}) U_\mu + 
    \varphi_{,\mu}+s\theta_{,\mu}+\beta_A\alpha^A_{,\mu} = 0 \,,
\end{align}
where $\upmu_{\rm tr}$ is the contribution of the nonminimal trace coupling to the chemical potential given by
\begin{align}
\label{chemical_potential_brown}
    \upmu_{\rm tr} = \frac{1}{2\kappa}f_T \frac{\partial T}{\partial n} \,.
\end{align}
The modifications of Eqs.~\eqref{seq} and~\eqref{Jeq} are due to matter tensor couplings which contribute to the temperature and chemical potential. Since the matter tensor of a fluid will always depend on the particle-number density, the term $\upmu_{\rm tr}$ will in general contribute to the fluid's equations of motion. 

\subsection{Conservation equations via field equations}
\label{secconsfield}

We now check the conservation equations for the effective energy-momentum tensor. Typically, the continuity equation gives $\nabla^\mu T^{\rm eff}_{\mu\nu}=0$. This is not true in general in the context of the coupling with between geometry and matter trace. We can easily see this by taking the divergence of the field equations~\eqref{eq_our_field_eqn}:
%%%%%%%%%%%%%%%%%%%%%%%%%%%%%%
\begin{align}
    \nabla^\mu(f_R R_{\mu\nu}) + (\nabla_\nu \Box - \Box\nabla_\nu) f_R - 
    \frac{1}{2} \nabla_\nu f &= 
    \nabla^\mu \Bigl(\kappa T_{\mu\nu} - \frac{1}{2}\ f_T n\frac{\partial T}{\partial n} 
    (g_{\mu\nu} + U_\mu U_\nu)\Bigr) \,, \\
    f_R \nabla^\mu R_{\mu\nu} - 
    \frac{1}{2} f_R \nabla_\nu R - 
    \frac{1}{2} f_T \nabla_\nu T &= 
    \nabla^\mu \Bigl(\kappa T_{\mu\nu} - \frac{1}{2}\ f_T n\frac{\partial T}{\partial n} 
    (g_{\mu\nu} + U_\mu U_\nu)\Bigr) \,, \\
    -\frac{1}{2} f_T \nabla_\nu T &= 
    \nabla^\mu \Bigl(\kappa T_{\mu\nu} - \frac{1}{2}\ f_T n\frac{\partial T}{\partial n} 
    (g_{\mu\nu} + U_\mu U_\nu)\Bigr) \,.
\end{align}
This implies that
\begin{align}
\label{eq_our_div}
    2\kappa \nabla^\mu T_{\mu\nu} = \nabla^\mu \Bigl(f_T n\frac{\partial T}{\partial n} (g_{\mu\nu} + U_\mu U_\nu)\Bigr) -
    f_T g_{\mu\nu} \nabla^\mu T \,,
\end{align}
which we can also write conveniently as
\begin{align}
    \label{eq_effective_divergence}
    \nabla^\mu T^{\rm eff}_{\mu\nu} = 
    -\frac{1}{2\kappa} f_T \nabla_\nu T \,.
\end{align}
While this conservation equation is valid in general, it will be convenient for us to consider its projection along the fluid's four-velocity and along the perpendicular directions. To do this, we define the usual projection $h^{\mu\nu}=U^\mu U^\nu+g^{\mu\nu}$.

First, contracting with $U^\nu$,
\begin{align}
    2\kappa U^\nu \nabla^\mu T_{\mu\nu} &= U^\nu(g_{\mu\nu} + U_\mu U_\nu) \nabla^\mu \Bigl(f_T n\frac{\partial T}{\partial n}\Bigr)  +
    U^\nu \Bigl(f_T n\frac{\partial T}{\partial n}\Bigr) \nabla^\mu(U_\mu U_\nu) -f_T U^\nu g_{\mu\nu} \nabla^\mu T \notag \\
     &= 
    -\Bigl(f_T n\frac{\partial T}{\partial n}\Bigr) \nabla^\mu U_\mu - f_T U_\mu \Bigl(\frac{\partial T}{\partial n}\nabla^\mu n + \frac{\partial T}{\partial s} \nabla^\mu s\Bigr) \,,
    \label{uconservation1}
\end{align}
%%%%%%%%%%%%%%%%%%%%%%%%%%%%%%
where we have used the relations $U^\mu h_{\mu\nu}=0$ and $U^\nu \nabla^\mu U_\nu=0$. Noticing also that $U_\mu \nabla^\mu s =0$ and $U_\mu \nabla^\mu n = -n \nabla^\mu U_\mu$, we have 
%%%%%%%%%%%%%%%%%%%%%%%%%%%%%%
\begin{align}
\label{uconservation2}
    2\kappa U^\nu \nabla^\mu T_{\mu\nu} &= 
    -\Bigl(f_T n\frac{\partial T}{\partial n}\Bigr) \nabla^\mu U_\mu + f_T  \Bigl(\frac{\partial T}{\partial n}n \nabla^\mu U_\mu \Bigr) = 0 \,.
\end{align}
%%%%%%%%%%%%%%%%%%%%%%%%%%%%%%
The perpendicular contraction is given by contracting Eq.~\eqref{eq_our_div}  with $h^{\nu\sigma}$ and rewriting it as
%%%%%%%%%%%%%%%%%%%%%%%%%%%%%%
\begin{align}
\label{eq_perpen_cons_field}
     2\kappa h^{\nu\sigma} \nabla^\mu T_{\mu\nu} &=   h^{\sigma}_\mu{} \nabla^\mu(f_T n\frac{\partial T}{\partial n}) -  f_T h^{\nu\sigma} \nabla_\nu T +(f_T n\frac{\partial T}{\partial n}) h^{\nu\sigma}\nabla^\mu h_{\mu\nu} \,.
\end{align}
%%%%%%%%%%%%%%%%%%%%%%%%%%%%%%
We have the identity $ h^{\nu\sigma}\nabla^\mu h_{\mu\nu} = U^\mu \nabla_\mu U^\sigma$. Applying the identity, with further manipulation of the right-hand side, gives the result
%%%%%%%%%%%%%%%%%%%%%%%%%%%%%%
\begin{align}
    2\kappa h^{\nu}_{\sigma} \nabla^\mu T_{\mu\nu} &= nh_{\sigma}^{\mu}\nabla_\mu (f_T \frac{\partial T}{\partial n}) -
     f_T \frac{\partial T}{\partial s} \nabla_\sigma s
      +  f_T n\frac{\partial T}{\partial n} U^\mu \nabla_\mu U_\sigma \,.
      \label{hconservation1a}
\end{align}
%%%%%%%%%%%%%%%%%%%%%%%%%%%%%%
Or using the identity for $U^\mu \nabla_\mu U_\sigma$ again, we can write the result nicely as
%%%%%%%%%%%%%%%%%%%%%%%%%%%%%%
\begin{align}
    2\kappa h^{\nu}_{\sigma} \nabla^\mu T_{\mu\nu} &=
    h_{\sigma}^{\nu} \Big[n \nabla_\mu (f_T \frac{\partial T}{\partial n} h^\mu_\nu) - f_T \frac{\partial T}{\partial s} \nabla_\nu s \Bigr] \,.
    \label{hconservation1}
\end{align}

We find that the perpendicular contraction of the conservation equations does not vanish. This is expected from the relation in Eq.~\eqref{eq_effective_divergence}. Since the parallel contraction vanished, this tells us that the entire contribution of the coupling with the matter trace in the conservation equations is in the perpendicular direction to the flow. 

\subsection{Conservation equations via fluid equation}

We will now rederive the the conservation equation from the fluid equation using only the equations of motion, without referring to the field equations. This will further verify the internal consistency of our equations and can be seen as the final step in demonstrating the correctness of our approach. 

By applying the covariant derivative to the energy-momentum tensor one simply gets $\nabla_\mu T^{\mu \nu} =\nabla_\mu (\rho U^\mu U^\nu )+\nabla_\mu(ph^{\mu\nu}) $. We will evaluate this along the flow and in the perpendicular direction. If we contract this equation with $U_\nu$, it is straightforward to verify that $U_\nu\nabla_\mu T^{\mu\nu}=0$ which matches the result of Eq.~\eqref{uconservation2}.

To evaluate $h_{\nu \sigma}\nabla_\mu T^{\mu \nu}$, we begin with [see Eq.~(2.13) in \cite{Brown_1993}]
\begin{align}
     h^{\nu}_{\sigma} \nabla^\mu T_{\mu\nu} =
     2nU^\mu\nabla_{[\mu}(\upmu U_{\sigma]}) - 
     h^\mu_\sigma \frac{\partial\rho}{\partial n}\nabla_\mu s\,,
     \label{hconservation2}
\end{align}
which is unaffected by the trace coupling. Additional details regarding this derivation can be found in Appendix~\ref{appen_cons_fluid}.

The key idea is to note that the term $\upmu U_{\sigma}$ is given by the fluid equation, Eq.~\eqref{Jeq}, which can thus be used to find an explicit expression for $\nabla_{[\mu}(\upmu U_{\sigma]})$. To start, we have
\begin{align}
    \nabla_\mu(\upmu U_\sigma) - \nabla_\mu\Bigl[\frac{f_T}{2\kappa}\Bigl(3 n\frac{\partial^2 \rho}{\partial n^2}
    - \frac{\partial \rho}{\partial n}\Bigr) U_\sigma\Bigr] + 
    \nabla_\mu \varphi_{,\sigma}+\nabla_\mu(s\theta_{,\sigma})+\nabla_\mu(\beta_A\alpha^A_{,\sigma}) = 0\,,
\end{align}
so that taking the skew-symmetric part yields
\begin{align}
    \nabla_{[\mu}(\upmu U_{\sigma]}) - \nabla_{[\mu}\Bigl[\frac{f_T}{2\kappa}\Bigl(3 n\frac{\partial^2 \rho}{\partial n^2}
    - \frac{\partial \rho}{\partial n}\Bigr) U_{\sigma]}\Bigr] + \nabla_{[\mu}\varphi_{,\sigma]}+\nabla_{[\mu}(s\theta_{,\sigma]})+\nabla_{[\mu}(\beta_A\alpha^A_{,\sigma]}) = 0 \,.
\end{align}
Using the fluid's equations of motion removes two out of the three final terms:
\begin{align}
    U^\mu \nabla_{[\mu}(\upmu U_{\sigma]}) - U^\mu \nabla_{[\mu}\Bigl[\frac{f_T}{2\kappa} \Bigl(3 n\frac{\partial^2 \rho}{\partial n^2}
    - \frac{\partial \rho}{\partial n}\Bigr) U_{\sigma]}\Bigr] + U^\mu\nabla_{[\mu}(s\theta_{,\sigma]}) = 0 \,.
\end{align}
One can now rewrite the final term, using again the fluid equations, to find 
\begin{align}
    2U^\mu\nabla_{[\mu}(s\theta_{,\sigma]}) = 
    -\nabla_{\sigma} s (U^\mu\nabla_\mu \theta)\,,
\end{align}
which can also be found in~\cite{Brown_1993} and is unaffected by the trace coupling. Finally, one uses Eq.~\eqref{seq}:
\begin{align}
    2U^\mu \nabla_{[\mu}(\upmu U_{\sigma]}) &= 2U^\mu \nabla_{[\mu}\Bigl[\frac{f_T}{2\kappa} \Bigl(3 n\frac{\partial^2 \rho}{\partial n^2}
    - \frac{\partial \rho}{\partial n}\Bigr) U_{\sigma]}\Bigr] -2U^\mu\nabla_{[\mu}(s\theta_{,\sigma]}) \notag \\ &=
    2U^\mu \nabla_{[\mu}\Bigl[\frac{f_T}{2\kappa} \Bigl(3 n\frac{\partial^2 \rho}{\partial n^2}
    - \frac{\partial \rho}{\partial n}\Bigr) U_{\sigma]}\Bigr] +
    \frac{1}{n}\Bigl\{\frac{\partial \rho}{\partial s} - \frac{f_T}{2\kappa} \Bigl(3 n\frac{\partial^2 \rho}{\partial n \partial s} - 4\frac{\partial \rho}{\partial s} \Bigr)\Bigr\}\nabla_\sigma s \,,
\end{align}
which means we now have an expression for the skew part. This can be substituted into Eq.~\eqref{hconservation2}, which after some calculations yields
\begin{align}
    h^{\nu}_{\sigma} \nabla^\mu T_{\mu\nu} =  
    2nU^\mu \nabla_{[\mu}\Bigl[\frac{f_T}{2\kappa} \Bigl(3 n\frac{\partial^2 \rho}{\partial n^2} - \frac{\partial \rho}{\partial n}\Bigr) U_{\sigma]}\Bigr] -
    \frac{f_T}{2\kappa}\frac{\partial T}{\partial s}\nabla_\sigma s \,.
\end{align}

Next, we can expand the skew-symmetrization and perform the covariant differentiation using the product rule, which gives three terms. These can be slightly rewritten and combined in such a way that one arrives at 
\begin{align}
\label{eq_perpen_fluid_cons}
    2\kappa h^{\nu}_{\sigma} \nabla^\mu T_{\mu\nu} &=
    h^\mu_{\alpha} n \nabla_{\mu}\Bigl[f_T \frac{\partial T}{\partial n} \Bigr] + n \Bigl[f_T \frac{\partial T}{\partial n}\Big] U^\mu \nabla_{\mu} U_{\sigma} -
    f_T\frac{\partial T}{\partial s}\nabla_\sigma s \notag \\
     &=
    h_{\sigma}^{\nu} \Big[n \nabla_\mu (f_T \frac{\partial T}{\partial n} h^\mu_\nu) - f_T \frac{\partial T}{\partial s} \nabla_\nu s \Bigr]
    \,,
\end{align}
which agrees with Eq.~\eqref{hconservation1}, as one would expect. This implies that the conservation equations derived from the fluid equations of motion agree with the conservation equations derived from the gravitational field equations.

\subsection{Effective equations and dark sector interactions}
\label{Brown_dark_sector}

The nonvanishing right-hand sides of the conservation equations~\eqref{hconservation1} and~\eqref{eq_perpen_fluid_cons} demonstrate the nonconservation of the energy-momentum tensor of the fluid, a common feature of theories with matter-curvature couplings. One can study the effects of such additional terms in the context of dark sector interactions by identifying suitable energy-momentum tensor contributions. To arrive at effective field equations, we begin with 
\begin{align}
    f(R,T) &= R+\bar{f}(R,T) \,,
\end{align}
with $f_T = \bar{f}_T$ and $f_R = 1 + \bar{f}_R$. This choice of function will isolate the Einstein tensor. The field equations~\eqref{eq_our_field_eqn} can now be written in a form similar to the Einstein field equations:
\begin{align}
    G_{\mu\nu} =  \kappa T_{\mu\nu} - \frac{1}{2} \bar{f}_T n\frac{\partial T}{\partial n} 
    (g_{\mu\nu} + U_\mu U_\nu)+ \frac{1}{2} \bar{f} g_{\mu\nu} - \bar{f}_R R_{\mu\nu} - \big(g_{\mu\nu}\Box - \nabla_\mu\nabla_\nu\big) \bar{f}_R\,,
    \label{eff_new1}
\end{align}
with all terms due to the nonminimal coupling moved onto the matter side of the equation. The first term on the right-hand side is the fluid's matter energy-momentum tensor. The remaining terms are matter and geometry dependent since $\bar{f}$ depends on the Ricci scalar and the matter tensor's trace. These remaining terms can be associated with dark energy and dark matter. We note, however, that this split is not unique. We set the dark energy part to be
\begin{align}
    T_{\mu\nu}^{\rm DE} :=  \frac{1}{2\kappa} \Bigl(\bar{f} - \bar{f}_T n\frac{\partial T}{\partial n}\Bigr)g_{\mu\nu} \,.
\end{align}
This is motivated by the form of the cosmological constant term $\Lambda g_{\mu\nu}$. Assuming that dark matter is pressureless, it is reasonable to define the following dark matter part:
\begin{align}
    T_{\mu\nu}^{\rm DM} :=  - \frac{1}{2\kappa} \bar{f}_T n\frac{\partial T}{\partial n} U_{\mu} U_{\nu} \,.
\end{align}
The remainder of Eq.~\eqref{eff_new1} is an additional matter component:
\begin{align}
    T_{\mu\nu}^{\bar{f}} := - \frac{1}{\kappa}\bigl(R_{\mu\nu} + g_{\mu\nu}\Box - \nabla_\mu\nabla_\nu\bigr) \bar{f}_R\,.
\end{align}
Using this notation, we can rewrite the field equations in a convenient form:
\begin{align}
    G_{\mu\nu} =  \kappa \Bigl(T_{\mu\nu} + T_{\mu\nu}^{\rm DM} + T_{\mu\nu}^{\rm DE} + T_{\mu\nu}^{\bar{f}}\Bigr) \,.
\end{align}
One can immediately show that
\begin{align}
    \nabla^\mu T_{\mu\nu}^{\bar{f}} = -\frac{1}{2\kappa}\bar{f}_{R}\nabla_\nu R =: -Q_\nu\,,
\end{align}
which can be interpreted as an interaction vector $Q_\nu$. For the other components, one cannot derive an independent nonconservation equation that only depends on that same interaction. Therefore, using the above notation, we can state that
\begin{align}
    \nabla^\mu \Bigl(T_{\mu\nu} + T_{\mu\nu}^{\rm DM} + T_{\mu\nu}^{\rm DE} \Bigr) =
    -\nabla^\mu T_{\mu\nu}^{\bar{f}} = Q_\nu \,.
\end{align}
Consequently, we have an interaction between the fluid, the identified dark energy and dark matter components. However, one cannot easily separate this interaction into the more standard form where $\nabla^\mu T_{\mu\nu}^{\rm DM} = -\nabla^\mu T_{\mu\nu}^{\rm DE}$, as discussed, for example in~\cite{Bansal:2024bbb, CarrilloGonzalez:2017cll}. This follows from our fluid equations~\eqref{alphaeq}-\eqref{phieq} that contain the geometrical couplings, which imply that $\nabla^\mu T_{\mu\nu} \neq 0$. Even absorbing $T_{\mu\nu}^{\bar{f}}$ into either dark energy or dark matter will not change this outcome.

\section{SCHUTZ'S VELOCITY POTENTIALS APPROACH}
\label{sec_Schutz}

In this section, we work with the relativistic perfect fluid Lagrangian given by Schutz~\cite{Schutz:1970}. We follow the structure of the previous section to compare between the two pictures and verify our approach of using the off-shell Lagrangian. Here, the Lagrangian is given simply by $\mathcal{L}_{\rm fluid}=p\sqrt{-g}$. The full description of the fluid is contained in the velocity potentials. The four-velocity unit vector is defined as a function of the independent variables with $U_\nu=\upmu^{-1}(\phi_{,\nu}+\alpha\beta_{,\nu}+\theta s_{,\nu})=\upmu^{-1}V_\nu$. The fluid action for the relativistic perfect fluid in this approach is
\begin{align}
\label{eq_Schutz_lagrangian}
    S_{\rm fluid} = \int p(\upmu,s)\sqrt{-g}\, d^4x\,,
\end{align}
with $\upmu=\sqrt{-g^{\mu\nu}V_\mu V_\nu}$. Thus, while $\upmu$ is a potential in this approach, it is not an independent variable. The independent variables which we vary the action with respect to are
\begin{align}
    \{g^{\mu\nu}, \alpha,  \beta, \phi, \theta, s\} \,.
\end{align}

Here, the density is a derived quantity, defined by $\rho = \upmu (\partial p/\partial\upmu) - p$. The Legendre transformation governs the thermodynamic equivalence of the choice of $\upmu$ as a parameter instead of $n$ as in Brown's fluid formalism. The trace of the perfect fluid energy-momentum tensor takes the form 
\begin{align}
\label{schutz_trace_matter}
    T=-\rho+3p = -\upmu\frac{\partial p}{\partial\upmu} + 4p \,.
\end{align}

We already notice here the analog between $\rho$ and $p$ in Brown's and Schutz's formalisms, a feature that will continue to emerge throughout the section. In a similar fashion, we start by deriving the field equations and equations of motion and show the on-shell reduction, thermodynamic quantities, and conservation equations.

\subsection{Gravitational field equations}
\label{sec_grav_field_eqns_schutz}
We start again by varying the action for $f(R,T)$ model, this time with a matter Lagrangian density given by Eq.~\eqref{eq_Schutz_lagrangian}. We use our previous results in Sec.~\ref{Brown Gravitational field equations} since the difference in this picture is only in the trace of the fluid matter and the Lagrangian. We have
\begin{align}
    \delta S = \frac{1}{2\kappa}\int &\Bigl[
    f_R R_{\mu\nu}
    +\big(g_{\mu\nu}\Box - \nabla_\mu\nabla_\nu\big) f_R - \frac{1}{2}fg_{\mu\nu} \Bigr]\delta g^{\mu\nu} \sqrt{-g}\,d^4x \notag \\
    &+\frac{1}{2\kappa} \int f_T \delta T \sqrt{-g}\,d^4x +\int\sqrt{-g} \big[\delta p-\frac{1}{2}pg_{\mu\nu}\delta g^{\mu\nu}\big] d^4x \,.
\end{align}
The variation of the energy-momentum trace $\delta T$ can be derived from Eq.~\eqref{schutz_trace_matter}: 
\begin{align}
\label{Schutz_T_variation}
    \delta T &= \Big( 3\frac{\partial p}{\partial \upmu}-\upmu \frac{\partial^2p}{\partial \upmu^2}\Big)\delta \upmu +\Big(4\frac{\partial p}{\partial s}-\upmu\frac{\partial^2 p}{\partial s \partial \upmu} \Big) \delta s = \frac{\partial T}{\partial \upmu} \delta \upmu+\frac{\partial T}{\partial s}\delta s \,,
\end{align}
which is again the result that we should get. Since $\upmu^2=-g^{\mu\nu}V_\mu V_\nu$, we can find the variation by using
\begin{align}
    2\upmu\delta \upmu&=-V_\mu V_\nu\delta g^{\mu\nu}-2g^{\mu\nu}V_\mu \delta V_\nu \,, \\
    \Rightarrow\delta \upmu &=-\frac{V_\mu V_\nu}{2\upmu}\delta g^{\mu\nu}-\frac{V^\nu}{\upmu}\Big[\delta \phi_{,\nu}+\beta_{,\nu}\delta \alpha+\alpha \delta \beta_{,\nu}+s_{,\nu}\delta \theta+\theta\delta s_{,\nu}     \Big] \label{Schutz_delta_upmu} \,.
\end{align}
We substitute the variation $\delta \upmu$ back in $\delta T$ to get the final expression in terms of independent variables:
\begin{align}
   \delta T    &=\frac{\partial T}{\partial \upmu}\bigg[-\frac{1}{2}\upmu U_\mu U_\nu\delta g^{\mu\nu} - U^\nu\Big(\delta \phi_{,\nu}+\beta_{,\nu}\delta \alpha+\alpha \delta \beta_{,\nu}+s_{,\nu}\delta \theta+\theta\delta s_{,\nu} \Big) \bigg] + \frac{\partial T}{\partial s}\delta s  \,.
\end{align}
Similarly, we write out $\delta p$ in terms of partial derivates of $\upmu$ and $s$ and use Eq.~\eqref{Schutz_delta_upmu} to get
\begin{align}
    \delta p &= \frac{\partial p}{\partial \upmu}\delta \upmu + \frac{\partial p}{\partial s}\delta s \notag \\
    &=-\frac{\partial p}{\partial \upmu} \bigg[\frac{1}{2}\upmu U_\mu U_\nu\delta g^{\mu\nu}+U^\nu\Big(\delta \phi_{,\nu}+\beta_{,\nu}\delta \alpha+\alpha \delta \beta_{,\nu}+s_{,\nu}\delta \theta+\theta\delta s_{,\nu} \Big) \bigg]  + \frac{\partial p}{\partial s}\delta s \,.
\end{align}
Now we collect all the terms and write the action variation in terms of the independent variables:
\begin{multline}
\label{Schutz_full_action}
    \delta S = \frac{1}{2\kappa}\int\sqrt{-g} \Bigg[ \Bigl(
    f_R R_{\mu\nu}
    +\big(g_{\mu\nu}\Box - \nabla_\mu\nabla_\nu\big) f_R - \frac{1}{2}fg_{\mu\nu} \Bigr)\delta g^{\mu\nu}\\ 
    + f_T \bigg[\frac{\partial T}{\partial \upmu}\bigg(-\frac{1}{2}\upmu U_\mu U_\nu\delta g^{\mu\nu} - U^\nu\Big(\delta \phi_{,\nu}+\beta_{,\nu}\delta \alpha+\alpha \delta \beta_{,\nu}+s_{,\nu}\delta \theta+\theta\delta s_{,\nu} \Big) \bigg)+\frac{\partial T}{\partial s}\delta s\bigg]
    \\+2\kappa\bigg[-\frac{\partial p}{\partial \upmu} \bigg(\frac{1}{2}\upmu U_\mu U_\nu\delta g^{\mu\nu}+U^\nu\Big(\delta \phi_{,\nu}+\beta_{,\nu}\delta \alpha+\alpha \delta \beta_{,\nu}+s_{,\nu}\delta \theta+\theta\delta s_{,\nu} \Big) \bigg)\\ + \frac{\partial p}{\partial s}\delta s
    -\frac{1}{2}pg_{\mu\nu}\delta g^{\mu\nu}\bigg]\Bigg] d^4x \,.
\end{multline}
We are ready now to retrieve the field equations and equations of motion from the action variation. First, the field equations can be read off from the $\delta g^{\mu\nu}$ terms and substituting the definition of $T_{\mu\nu}$ to get
\begin{align}
\label{eq_Schutz_field_eqn}
    f_R R_{\mu\nu}
    +\big(g_{\mu\nu}\Box - \nabla_\mu\nabla_\nu\big) f_R - \frac{1}{2}fg_{\mu\nu} &=\kappa T_{\mu\nu} +\frac{1}{2}f_T  \upmu\frac{\partial T}{\partial \upmu}U_\mu U_\nu \,.
\end{align}
On the right-hand side, we get an extra term that depends on the partial derivate of the function $f(R,T)$ with respect to $T$ and the derivative of the matter trace with respect to the chemical potential $\upmu$. This again suggests an effective energy-momentum tensor given by
\begin{align}
\label{Schutz_Teff}
    T_{\mu\nu}^{\rm eff} = T_{\mu\nu} + \frac{f_T}{2\kappa}
    \upmu\frac{\partial T}{\partial \upmu} U_\mu U_\nu \,.
\end{align}
Notice that here the extra term couples to four-velocity unit vectors $U_\mu U_\nu$ but not the metric, which, by using the definition of $T_{\mu\nu}$, suggests an effective energy density rather than pressure, given by 
\begin{align}
    \rho_{\rm eff} := 
    \rho + \frac{f_T}{2\kappa} \upmu\frac{\partial T}{\partial \upmu} \,.
    \label{rhoeffSchutz}
\end{align}
Not surprisingly, we get analogous results to that found in the Brown picture. Here, we see the exchange of the derived quantity between $p$ in Brown's picture and $\rho$ in Schutz's picture led to the exchange of their effective role in the field equations. The expression for $\rho_{\rm eff}$ agrees with the on-shell expression in the argument presented in Sec.~\ref{sec:onshell}. Of course, one has to be careful when changing the variables $n$ and $\upmu$; however, it should be clear at this point that the trace coupling can only yield a contribution that affects the energy density or the pressure.

\subsection{Equations of motion}

Before discussing the field equations and the effective quantities further, we derive the fluid's equations of motion from the remaining variations in Eq.~\eqref{Schutz_full_action}. By collecting the terms for each of the five potential variations we get
\begin{alignat}{2}
    \delta\alpha:& \qquad & \Big(\frac{f_T}{2 \kappa}\frac{\partial T}{\partial \upmu}+\frac{\partial p}{\partial \upmu}\Big)\big(U^\nu \beta_{,\nu} \big) &= 0 \,, \label{delta_alpha} \\
    \delta\beta:& \qquad &  \partial_\nu\bigg[\Big(\frac{f_T}{2 \kappa}\frac{\partial T}{\partial \upmu}+\frac{\partial p}{\partial \upmu}\Big)\alpha U^\nu\bigg] &= 0 \,, \label{delta_beta} \\
    \delta\theta:& \qquad &\Big(\frac{f_T}{2 \kappa}\frac{\partial T}{\partial \upmu}+\frac{\partial p}{\partial \upmu}\Big)\big(U^\nu s_{,\nu} \big) &= 0 \,, \label{delta_theta} \\
    \delta\phi:& \qquad & \partial_\nu\bigg[\Big(\frac{f_T}{2 \kappa}\frac{\partial T}{\partial \upmu}+\frac{\partial p}{\partial \upmu}\Big) U^\nu\bigg] &= 0 \,,  \label{delta_phi}\\
    \delta s:& \qquad & \partial_\nu\bigg[\Big(\frac{f_T}{2 \kappa}\frac{\partial T}{\partial \upmu}+\frac{\partial p}{\partial \upmu}\Big)\theta U^\nu\bigg] + \frac{\partial p}{\partial s} 
    + \frac{1}{2 \kappa} f_T \frac{\partial T}{\partial s}  &= 0 \label{delta_s} \,.
\end{alignat} 
We can immediately see from Eqs.~\eqref{delta_alpha} and~\eqref{delta_theta} that $U^\nu \beta_{,\nu}=0$ and $U^\nu s_{,\nu}=0$. Further, by comparing these equations with the ones obtained by Schutz~\cite{Schutz:1970}, we are tempted to redefine the parameter $n$ from $n=\partial p/\partial \upmu$ to be
\begin{align}
\label{Schutz_n}
    n \equiv \frac{\partial p}{\partial \upmu}+\frac{f_T}{2 \kappa}\frac{\partial T}{\partial \upmu} \,;
\end{align}
then we find from Eq.~\eqref{delta_phi} that $\partial_\nu \big(nU^\nu\big)=0$. Using this relation and the product rule in Eqs.~\eqref{delta_beta} and~\eqref{delta_s} allows us to write the equations of motion as
\begin{alignat}{2}
    \delta\alpha:& \qquad & U^\nu \beta_{,\nu} &= 0  \,, \qquad \label{Schutz_delta_alpha}\\
    \delta\beta:& \qquad &  U^\nu \alpha_{,\nu}  &= 0 \,, \qquad\label{Schutz_delta_beta}\\
    \delta\theta:& \qquad &U^\nu s_{,\nu}&= 0 \,, \qquad \label{Schutz_delta_theta} \\
    \delta\phi:& \qquad & \partial_\nu\big(n U^\nu \big) &= 0 \,, \qquad \label{Schutz_delta_phi}\\
    \delta s:& \qquad & n U^\nu \theta_{,\nu}+\frac{\partial p}{\partial s}+\frac{f_T}{2 \kappa} \frac{\partial T}{\partial s} &= 0 \,. \label{Schutz_delta_s}
\end{alignat}
The equations of motion~\eqref{Schutz_delta_alpha}-\eqref{Schutz_delta_phi} now match the expected result found by Schutz \cite{Schutz:1970} with the new definition of $n$ in Eq.~\eqref{Schutz_n}. We see, however, that the entropy variation $\delta s$ leads to an additional term $(2\kappa)^{-1} f_T \partial T/\partial s$ which is a result of the nonminimal coupling between curvature and the matter trace. In fact, even the factor $n$  in Eq.~\eqref{Schutz_delta_phi} which ensures the conservation of the particle number now depends on the nonminimal coupling between $R$ and $T$. Those effects will be discussed in more detail in Secs.~\ref{Schutz Thermodynamic quantities} and~\ref{sec_implications}.

\subsection{On-shell fluid Lagrangian}

We want to make sure again that we retrieve the the on-shell form of the Lagrangian when the equations of motion are taken into consideration. We already have the off-shell Lagrangian as
\begin{align}
    S_{\textrm{fluid}} = \int p(\mu,s) \sqrt{-g}\,d^4x \qquad \textrm{off shell} \,,
\end{align}
in the velocity potentials picture. But we see that if we start with this expression, use $p=\upmu\partial p/\partial \upmu - \rho$, and then substitute Eq.~\eqref{Schutz_n} for $\partial p/\partial \upmu$, we get
\begin{align}
    S_{\textrm{fluid}} &= \int p \sqrt{-g}\,d^4x= \int \Big(\upmu\frac{\partial p}{\partial \upmu} -\rho \Big) \sqrt{-g}\,d^4x  \notag \\
    &= \int\Big(\upmu \big(n-\frac{f_T}{2\kappa}\frac{\partial T}{\partial \upmu} \big)-\rho\Big)\sqrt{-g}\,d^4x \notag \\
    &=\int \upmu n \sqrt{-g}\,d^4x-\int\big(\rho+\frac{f_T}{2\kappa}\upmu\frac{\partial T}{\partial \upmu}\big)\sqrt{-g}\,d^4x = \int \upmu n \sqrt{-g}\,d^4x-\int\rho_{\textrm{eff}}\sqrt{-g}\,d^4x \label{Schutz_action_onshell_step1}\,,
\end{align}
where we used the definition of $\rho_{\textrm{eff}}$ from Eq.~\eqref{rhoeffSchutz}.

Now, it can be shown from the definition of $U_\nu$ and using the equations of motion that 
\begin{align}
    U^\nu U_\nu &= U^\nu \upmu^{-1}(\phi_{,\nu}+\alpha \beta_{,\nu}+\theta s_{,\nu})=-1 \,, \\
    \Rightarrow -\upmu &=U^\nu \phi_{,\nu} \,.
\end{align}
where we used Eqs.~\eqref{Schutz_delta_alpha} and~\eqref{Schutz_delta_theta}. By substituting the expression for $\upmu$ back in the action and using the product rule of derivatives we find
\begin{align}
    \int \upmu n \sqrt{-g}\,d^4x &= -\int nU^\nu \phi_{,\nu}\sqrt{-g}\,d^4x
    =-\int\Big(\big(nU^\nu\phi  \big)_{,\nu}-\phi\big(nU^\nu\big)_{,\nu}\Big) \sqrt{-g}\,d^4x \,.
\end{align}
Now, we use Eq.~\eqref{Schutz_delta_phi} to eliminate the second term and we are left with a total derivative
\begin{align}
   \int \upmu n \sqrt{-g}\,d^4x &= -\int\big(U^\nu\phi n \big)_{,\nu} \sqrt{-g}\,d^4x 
    = - \int\big(U^\nu\phi n\sqrt{-g} \big)_{,\nu} d^4x = 0 \,,
\end{align}
which vanishes under action variation. Thus, going back to the action in Eq.~\eqref{Schutz_action_onshell_step1}, we end up with
\begin{align}
    S_{\textrm{fluid}}&= -\int\rho_{\textrm{eff}}\sqrt{-g}\,d^4x  \qquad  \textrm{on shell} \,.
\end{align}
Therefore, on shell, we do get back the expected expression after utilizing the equations of motion. This not only ensures the consistency of our results so far, but also strengthens the argument that one indeed has to be careful when working with the on-shell expression of the Lagrangian as it only holds up when the equations of motion are considered. However, when applying a variation principle, one must resort to the off-shell Lagrangian to get the full expression Eq.~\eqref{Schutz_full_action}.

\subsection{Thermodynamic quantities}
\label{Schutz Thermodynamic quantities}

The coupling with the matter trace in Brown's picture led to a modified expression for the chemical potential $\upmu$, which we called $\upmu_{\rm tr}$. The thermodynamics relations infer that there should be an analogous quantity in Schutz's picture related to the number density $n$. We already saw that the matter trace coupling led to modified expression presented in Eq.~\eqref{Schutz_n}. We now introduce the thermodynamic quantity $n_{\rm tr}$ where
\begin{align}
    n_{\rm{tr}} &= \frac{f_T}{2 \kappa}\frac{\partial T}{\partial \upmu} \qquad \textrm{and} \qquad \frac{\partial p}{\partial \upmu}  =n - n_{\rm{tr}} \,.
\end{align}
We can find a deeper meaning for the quantity $n_{\rm tr}$ by looking at the first law of thermodynamics where
\begin{align}
    dp = nd\upmu - n\mathcal{T}ds \,;
\end{align}
we see that letting the entropy change as a function of the chemical potential allows us to write
\begin{align}
    \frac{dp}{d\upmu} = n - n\mathcal{T}\frac{ds}{d\upmu} \,,
\end{align}
which allows us to identify $n_{\rm tr}$ with an entropy contribution that emerges from the coupling with the matter trace. The entropy contribution can be further seen by considering Eq.~\eqref{Schutz_delta_s}, where by using $\partial p/\partial s=-n\mathcal{T}$ from the the first law, we have
\begin{align}
    nU^\nu \theta_{,\nu} &- n\mathcal{T}+\frac{f_T}{2\kappa}\frac{\partial T}{\partial s}=0 \notag \,,\\
    U^\nu \theta_{,\nu} &= \mathcal{T}- \mathcal{T}_{\rm tr} \,.
\end{align}

Here we used the definition for $\mathcal{T}_{\rm tr}$ given in Eq.~\eqref{eq_temp_brown} and we get the expected result we found in the Brown picture. We see again that the contribution to the temperature arises from a geometric contribution due to the matter-geometry coupling in $f(R,T)$.

\subsection{Conservation equations via field equations}
\label{Schutz Conservation equations via field equations}

We again check the consistency of our approach by first deriving the conservation equations from the field equations in this section and then deriving them once more from the fluid equation in the following section.

We derived Eq.~\eqref{eq_effective_divergence} by taking the divergence of the field equations. We now use the velocity potentials expression for $T_{\mu\nu}^{\rm eff}$ given by Eq.~\eqref{Schutz_Teff}. As before, we take the contraction in the fluid direction $U^\nu$ and in the parallel direction $h^\nu_\sigma$. Starting with the parallel contraction
\begin{align}
    U^\nu\nabla^\mu T_{\mu\nu}^{\rm{eff}} &= -\frac{f_T}{2\kappa} U^\nu\nabla_\nu T 
    = -\frac{f_T}{2\kappa}\frac{\partial T}{\partial \upmu} U^\nu\nabla_\nu \upmu -\frac{f_T}{2\kappa}\frac{\partial T}{\partial s} U^\nu\nabla_\nu s
    = -\frac{f_T}{2\kappa}\frac{\partial T}{\partial \upmu} U^\nu\nabla_\nu \upmu \label{eq_teff_rhs}\,,
\end{align}
where the contracted derivative of the entropy term is eliminated by the equation of motion~\eqref{Schutz_delta_theta}. By substituting the expression for $T_{\mu\nu}^{\rm eff}$ from Eq.~\eqref{Schutz_Teff} we get
\begin{align}
    U^\nu\nabla^\mu T_{\mu\nu}^{\rm{eff}} &= U^\nu\nabla^\mu T_{\mu\nu} + U^\nu\nabla^\mu \Bigl(\frac{f_T}{2\kappa}
    \upmu\frac{\partial T}{\partial \upmu} U_\mu U_\nu\Bigr) \notag \\
    &= U^\nu\nabla^\mu T_{\mu\nu} -\upmu\nabla^\mu \Bigl(\frac{f_T}{2\kappa}
    \frac{\partial T}{\partial \upmu} U_\mu\Bigr)-\frac{f_T}{2\kappa}
    \frac{\partial T}{\partial \upmu} U_\mu\nabla^\mu \upmu \,,
\end{align}
where we used the identities $U^\nu U_\nu = -1$ and $U^\nu\nabla^\mu U_\nu=0$. Combining the two results for $U^\nu\nabla^\mu T_{\mu\nu}^{\rm{eff}}$ shows that
\begin{align}
    U^\nu\nabla^\mu T_{\mu\nu} &= \upmu\nabla^\mu \Bigl(\frac{f_T}{2\kappa}
    \frac{\partial T}{\partial \upmu} U_\mu\Bigr) \,,
\end{align}
which can be further simplified using Eqs.~\eqref{Schutz_n}  and~\eqref{Schutz_delta_phi} to get
\begin{align}
    U^\nu\nabla^\mu T_{\mu\nu} &= -\upmu\nabla^\mu \Big(\frac{\partial p}{\partial \upmu}U_\mu \Big) \,. \label{Schutz_parallel_field}
\end{align}
We already see that this result is different from the result obtained in Brown's Lagrange multipliers approach. The parallel contraction vanishes there due to the extra factor in the Brown's $T_{\mu\nu}^{\rm eff}$ being proportional to $h_{\mu\nu}$ while in the velocity potentials' $T_{\mu\nu}^{\rm eff}$ it is proportional to $U_\mu U_\nu$. Also, the reader of~\cite{Schutz:1970} notices that this term vanishes in the absence of couplings since $\partial p/\partial \upmu=n$ which would satisfy the equation of motion. However, the coupling between geometry and matter trace in $f(R,T)$ gravity introduces exactly the extra factor that contributes to the conservation equations parallel to the flow lines.

Now, we contract in the perpendicular direction to the flow. We want to show that $h^\nu_\sigma\nabla^\mu T_{\mu\nu}^{\rm eff}=-(2\kappa)^{-1}f_Th^\nu_\sigma\nabla_\nu T$. By substituting the expression for $T_{\mu\nu}^{\rm eff}$, we have
\begin{align}
     h^\nu_\sigma\nabla^\mu T_{\mu\nu} &= -h^\nu_\sigma\nabla^\mu \Big(\frac{f_T}{2\kappa}\upmu\frac{\partial T}{\partial \upmu}U_\mu U_\nu \Big)- \frac{f_T}{2\kappa} h^\nu_\sigma\nabla_\nu T = -h^\nu_\sigma \frac{f_T}{2\kappa}\upmu\frac{\partial T}{\partial \upmu}U_\mu\nabla^\mu  U_\nu- \frac{f_T}{2\kappa} h^\nu_\sigma\nabla_\nu T \,,
\end{align}
where again we used the identity $h^\nu_\sigma U_\nu=0$. By expanding the last term  we find
\begin{align}
    h^\nu_\sigma\nabla^\mu T_{\mu\nu}&= -\frac{f_T}{2\kappa}h^\nu_\sigma  \Big[ \upmu\frac{\partial T}{\partial \upmu}U_\mu\nabla^\mu  U_\nu+\frac{\partial T}{\partial \upmu}\nabla_\nu \upmu +\frac{\partial T}{\partial s}\nabla_\nu s\Big] \,.
\end{align}
Using the identities $h^\nu_\sigma h_{\mu\nu}=h^\nu_\sigma g_{\mu\nu}$ and $h^\nu_\sigma \nabla^\mu h_{\mu\nu}=h^\nu_\sigma U_\mu \nabla^\mu U_\nu$ allows us to combine the first two terms. After multiplying by $2\kappa$, the expression can be written in a form similar to that in Eq.~\eqref{hconservation1}:
\begin{align}
\label{Schutz_perpend_field_equation}
    2\kappa h^\nu_\sigma\nabla^\mu T_{\mu\nu} = -h^\nu_\sigma  \Big[f_T \frac{\partial T}{\partial \upmu} \nabla^\mu \big( h_{\mu\nu}\upmu \big)+f_T\frac{\partial T}{\partial s}\nabla_\nu s\Big] \,.
\end{align}
We have obtained the two expressions for the conservation equation parallel to the flow [Eq.~\eqref{Schutz_parallel_field}] and perpendicular to the flow [Eq.~\eqref{Schutz_perpend_field_equation}]. Now, we attempt to rederive the same expressions using the fluid equation without resorting to the field equations. 

\subsection{Conservation equations via fluid equations}

We start working from the fluid equation for $T_{\mu\nu}$ where in this picture we have $\rho+p=\upmu \partial p/\partial \upmu$. Then the divergence is
\begin{align}
    \nabla^\mu T_{\mu\nu}&=\nabla^\mu\Big( \upmu\frac{\partial p}{\partial \upmu} U_\mu U_\nu\Big) + g_{\mu\nu} \nabla^\mu p \,,
\end{align}
and just like before we check the contraction in the parallel and perpendicular directions. The contraction parallel to the flow is trivial where
\begin{align}
    U^\nu\nabla^\mu T_{\mu\nu}&=U^\nu\nabla^\mu\Big( \upmu\frac{\partial p}{\partial \upmu} U_\mu U_\nu\Big) + U^\nu g_{\mu\nu} \nabla^\mu p  =-\nabla^\mu\Big( \upmu\frac{\partial p}{\partial \upmu} U_\mu \Big) + U_\mu \nabla^\mu p \,.
\end{align}
By expanding the derivative of the pressure and using Eq.~\eqref{Schutz_delta_theta},
\begin{align}
   U^\nu\nabla^\mu T_{\mu\nu} &=-\nabla^\mu\Big( \upmu\frac{\partial p}{\partial \upmu} U_\mu \Big) +  U_\mu \frac{\partial p}{\partial \upmu} \nabla^\mu \upmu =-\upmu\nabla^\mu\Bigl(\frac{\partial p}{\partial \upmu} U_\mu \Bigr) \,,
\end{align}
which matches the result from the field equations in Eq.~\eqref{Schutz_parallel_field}, verifying our previous result that in this picture, the parallel contraction of the energy-momentum divergence does not vanish.

Now, we contract with $h^\nu_\sigma$ to check the conservation equations in the perpendicular direction:
\begin{align}
    h^\nu_\sigma \nabla^\mu T_{\mu\nu}&=h^\nu_\sigma\Big[\nabla^\mu\Big( \upmu\frac{\partial p}{\partial \upmu} U_\mu U_\nu\Big) + g_{\mu\nu} \nabla^\mu p\Big] \notag \\
    &=h^\nu_\sigma\Big[\upmu\frac{\partial p}{\partial \upmu} U_\mu\nabla^\mu  U_\nu + \frac{\partial p}{\partial \upmu} g_{\mu\nu} \nabla^\mu \upmu+ \frac{\partial p}{\partial s} \nabla_\nu s \Big]=h^\nu_\sigma\Big[\frac{\partial p}{\partial \upmu} \nabla^\mu \big( h_{\mu\nu}\upmu \big) + \frac{\partial p}{\partial s} \nabla_\nu s \Big] \,.
\end{align}
We can replace the partial derivatives of the pressure with partial derivatives of the mater trace by invoking the equations of motion Eqs.~\eqref{Schutz_n} and~\eqref{Schutz_delta_s}:
\begin{align}
    h^\nu_\sigma \nabla^\mu T_{\mu\nu} &=h^\nu_\sigma\Big[ \big( n -\frac{f_T}{2\kappa}\frac{\partial T}{\partial \upmu}\big) \nabla^\mu \big( h_{\mu\nu}\upmu \big)  - \big( nU^\lambda\theta_{,\lambda}+\frac{f_T}{2\kappa}\frac{\partial T}{\partial s} \big) \nabla_\nu s \Big] \notag \\
    &= h^\nu_\sigma\Big[  n\Big(h_{\mu\nu}\nabla^\mu \upmu + \upmu \nabla^\mu h_{\mu\nu}-U^\lambda\theta_{,\lambda}\nabla_\nu s\Big) -\frac{f_T}{2\kappa}\frac{\partial T}{\partial \upmu} \nabla^\mu \big( h_{\mu\nu}\upmu \big)  - \frac{f_T}{2\kappa}\frac{\partial T}{\partial s} \nabla_\nu s \Big] \,.
\end{align}
The term in round brackets vanishes [check Eq.~\eqref{special_appen_relation_2} in Appendix~\ref{appendix_vanishing_relation}], leaving us with 
\begin{align}
    2\kappa h^\nu_\sigma \nabla^\mu T_{\mu\nu} =& -h^\nu_\sigma\Big[f_T\frac{\partial T}{\partial \upmu} \nabla^\mu \big( h_{\mu\nu}\upmu \big)  + f_T\frac{\partial T}{\partial s} \nabla_\nu s \Big] \,,
\end{align}
which matches the result of Eq.~\eqref{Schutz_perpend_field_equation}. Thus, we have shown that our approach varying the action with the full fluid Lagrangian is also correct in Schutz's velocity potentials picture. 

With that we have strong confidence that the correct approach to vary the fluid action is to use the full off-shell fluid Lagrangian. We showed that this can be done in both Brown's approach with Lagrange multipliers or Schutz's approach based on velocity potentials. The results in both cases are consistent and lead to the same conclusions.

\subsection{Effective equations and  dark sector interactions}

We reexamine the mapping of our results to dark sector interactions. As before, we begin with $f(R,T) = R+\bar{f}(R,T)$ and find the field equation
\begin{align}
    G_{\mu\nu} =  \kappa T_{\mu\nu} + \frac{1}{2} \bar{f}_T \upmu\frac{\partial T}{\partial \upmu} U_\mu U_\nu + \frac{1}{2} \bar{f} g_{\mu\nu} - 
    \big(R_{\mu\nu} + g_{\mu\nu}\Box - \nabla_\mu\nabla_\nu\big) \bar{f}_R
    \label{eff_new2}
\end{align}
and now follow a slightly different approach to before. The right-hand side contains an effective energy density term, unlike the Brown approach where the extra term contributed to the pressure. This means that the Schutz approach yields an extra contribution which is pressureless. Let us therefore write 
\begin{align}
    G_{\mu\nu} &=  \kappa (T^{\rm eff}_{\mu\nu} + \hat{T}_{\mu\nu}^{\bar{f}}) \,,
\end{align}
where $\hat{T}_{\mu\nu}^{\bar{f}}$ contains all other terms on the right-hand side of Eq.~\eqref{eff_new2}. Here, the effective energy-momentum tensor is as in Eq.~\eqref{Schutz_Teff}, which means it contains the standard fluid plus a dark matter contribution. This formulation allows us to conclude that
\begin{align}
    \nabla^\mu T^{\rm eff}_{\mu\nu} = - 
    \nabla^\mu T_{\mu\nu}^{\bar{f}} = - \frac{1}{2\kappa}\bar{f}_T \nabla_\nu T =: \mathcal{Q}_\nu\,.
\end{align}
It would now be natural to interpret this as follows: The effective energy-momentum tensor, which contains the fluid and a dark matter contribution, is not conserved. It leads to an energy transfer to $T_{\mu\nu}^{\bar{f}}$ which we would then interpret as the dark energy component.

The identification of the various terms with either dark matter or dark energy is not unique as we have mentioned before. However, using Schutz's approach, one arrives at an effective energy density which lends itself to be interpreted as dark matter. In the other approach, see Sec.~\ref{Brown_dark_sector}, this identification is not as direct. We would therefore conclude that the fluid Lagrangian based on the pressure; this means the approach by Schutz is preferable for a nonminimally coupled dark matter model.

\section{IMPLICATIONS}
\label{sec_implications}

In the following, we will state the main implications of our work. From these implications, one finds that a vast amount of work done in $f(R,T)$ gravity needs to be revisited or must be considered incorrect. The latter applies to models linear in the trace of a perfect fluid energy-momentum tensor or models of the form $f(R,T)=f^{(1)}(R)+f^{(2)}(T)$. Moreover, we provide a simple argument why models linear in $T$ (or a perfect fluid) should not have been considered in the first place as such models cannot yield new physics.

\subsection{Comparison with Harko \textit{et al.}~\cite{Harko:2011kv}}

In \cite{Harko:2011kv}, Harko \textit{et al.} introduced $f(R,T)$ gravity in which curvature is coupled to the trace of the energy-momentum tensor of matter for several matter candidates, including a relativistic perfect fluid. We have shown that in such cases one has to ensure using the proper matter Lagrangian to derive the field equations and equations of motion as resorting to the on-shell Lagrangian does not give the full picture. Thus, we start our investigation of the implications of our results with comparison between the results in \cite{Harko:2011kv}.

 Using the field equation [see Eq.~(11) in~\cite{Harko:2011kv}] together with Eqs.~(20) and~(21) in~\cite{Harko:2011kv} one arrives at
\begin{align}
    \label{Harko_field_eqn}
    f_R R_{\mu\nu} + \big(g_{\mu\nu}\Box - \nabla_\mu\nabla_\nu\big) f_R - 
    \frac{1}{2} f g_{\mu\nu} = 
    \kappa T_{\mu\nu} + f_T (\rho+p) U_\mu U_\nu \,.
\end{align}
This field equation is similar, yet different from our result using the velocity potential approach in Eq.~\eqref{eq_Schutz_field_eqn} since
\begin{align}
    \frac{1}{2}\upmu\frac{\partial T}{\partial \upmu} = \frac{1}{2}\Bigl(3\upmu\frac{\partial p}{\partial \upmu} -\upmu^2\frac{\partial^2p}{\partial\upmu^2}\Bigr) \neq \rho + p \,,
\end{align}
in general. However, it is possible to provide a partial fix to the situation assuming fluids with constant entropy. In this case, let us choose an equation of state $p(\upmu)$ which we assume to be invertible so that we can find $\upmu=\upmu(p)$. Since the expression
\begin{align}
    \frac{1}{2}\Bigl(3\upmu\frac{\partial p}{\partial \upmu} -\upmu^2\frac{\partial^2p}{\partial\upmu^2}\Bigr)
\end{align}
is a function of $\upmu$, we can use the inverse equation of state to reexpress this as a function of the pressure $p$. This means, for given equation of state $p(\upmu)$ we can construct the function
\begin{align}
    \mathcal{F}(p) = \frac{1}{2}\Bigl(3\upmu\frac{\partial p}{\partial \upmu} -\upmu^2\frac{\partial^2p}{\partial\upmu^2}\Bigr)\biggr|_{\upmu=\upmu(p)}\,.
\end{align}
Let us now consider the energy density and pressure in Eq.~\eqref{Harko_field_eqn} and assume there exists an equation of state $\rho=\rho(p)$. If this equation of state satisfies
\begin{align}
    \mathcal{F}(p) = \rho(p) + p \,,
\end{align}
one would be able to reinterpret the results using a new equation of state. The caveat of this approach is that it is not clear whether the equation of state needed would be physically meaningful. It so happens that the situation is easier for linear equations of state and we can show explicitly how to map the original equations to the corrected field equations. 

\subsection{Models of the form $f(R,T)=f^{(1)}(R)+f^{(2)}(T)$}
\label{sec:summodels}

The important question to ask now is how do the modification inspired by considering the full fluid Lagrangians affect $f(R,T)$ gravity models. Let us substitute the ansatz $f(R,T)=f^{(1)}(R)+f^{(2)}(T)$ into the gravitational field equations~\eqref{eq_our_field_eqn} [a similar argument can be made for Eq.~\eqref{eq_Schutz_field_eqn}] which yields
\begin{align}
    \label{eq_our_field_eqn3}
    f^{(1)}_R R_{\mu\nu} + \big(g_{\mu\nu}\Box - \nabla_\mu\nabla_\nu\big) f^{(1)}_R - 
    \frac{1}{2} f^{(1)} g_{\mu\nu} - \frac{1}{2} f^{(2)} g_{\mu\nu} = 
    \kappa T_{\mu\nu} - \frac{1}{2}\ f^{(2)}_T n\frac{\partial T}{\partial n} 
    (g_{\mu\nu} + U_\mu U_\nu) \,.
\end{align}
We can now move the term $\frac{1}{2} f^{(2)} g_{\mu\nu}$ to the right-hand side and note that the new left-hand side is equivalent to standard $f(R)$ field equations. We call the new right-hand side the energy-momentum tensor of this model $\widetilde{T}_{\mu\nu}$ which means we have
\begin{align}
    \label{eq_our_field_eqn4}
    f^{(1)}_R R_{\mu\nu} + \big(g_{\mu\nu}\Box - \nabla_\mu\nabla_\nu\big) f^{(1)}_R - 
    \frac{1}{2} f^{(1)} g_{\mu\nu} = \kappa\widetilde{T}_{\mu\nu}\,.
\end{align}
This is nothing but $f(R)$ gravity with an usually constructed matter source; see~\cite{Fisher:2019ekh}. No background cosmological observation could ever distinguish this model from $f(R)$ gravity. One might argue that perturbation theory can come to the rescue; however, this is not the case. Any perturbations in the quantities $\tilde{\rho}$ and $\tilde{p}$ could be reexpressed in the original quantities, with no new information added. 

\subsection{Models linear in $T$ -- A heuristic argument}
\label{sec_models_linear_T}

The previous section shows that no model linear in $T$ is distinguishable from $f(R)$ gravity. However, this is indeed expected from first principles, whether one uses Brown's fluid approach or Schutz's approach (or any approach for that matter!). That is because the trace of the  perfect fluid energy-momentum tensor in the action is given $T\sqrt{-g}=(-\rho+3p)\sqrt{-g}$, which is thus proportional to $-\rho\sqrt{g}$ and  $p\sqrt{-g}$. But these are the expressions for the on-shell Lagrangian of a perfect fluid regardless of the full fluid Lagrangian expression. Consequently, any model using a linear trace coupling on shell must be equivalent to a perfect fluid with rescaled pressure or energy density. These are precisely our results of Eqs.~\eqref{eq_p_eff} and~\eqref{rhoeffSchutz}. Say we start with a model of the form given in Sec.~\ref{sec:summodels} which is linear in $T$. Setting $f(R,T)=f^{(1)}(R)+\chi T$ in Eq.~\eqref{eq_p_eff} yields
\begin{align}
    p_{\rm eff} = p - \frac{1}{2\kappa} \chi n\frac{\partial T}{\partial n} \,,
\end{align}
where $\chi$ is some proportionality constant. One can check that indeed such a model will merely affect the equation of state and does not contain any new physics. We show this explicitly by working, for example, with $\rho=n^{1+w}$, which is well known to give the linear equation of state $p=w\rho$. This follows directly from $n\partial \rho/\partial n=(1+w)\rho$ and the definition of pressure. In this case $T=-\rho+3p = (-1+3w)\rho$ and
\begin{align}
    n\frac{\partial T}{\partial n} = (3w-1)(1+w)\rho\,,
\end{align}
and we arrive at
\begin{align}
    p_{\rm eff} &= \tilde{w} \rho \,,
    \label{w1} \qquad  \textrm{where} \quad  \tilde{w} := w - \frac{1}{2\kappa} \chi (3w-1)(1+w) \,.
\end{align}
regardless of the complexity of the term $\tilde{w}$; all it did is to simply redefine the equation of state parameter with the same proportionality constant. Likewise, following Schutz's approach, we write $p=\upmu^{1+1/w}$, which allows us to get the linear equation of state as $\rho=p/w$ from $\rho+p=\upmu\partial p/\partial \upmu$. Then
\begin{align}
    \upmu\frac{\partial T}{\partial \upmu} = \frac{1}{w}(3w-1)(1+w)p \,.
\end{align}
Similarly, we substitute this expression in Eq.~\eqref{rhoeffSchutz} to get the effective energy density, which becomes
\begin{align}
    \rho_{\rm eff} = \rho+\frac{1}{2\kappa}\chi\upmu \frac{\partial T}{\partial \upmu} &=
    \frac{p}{w} + \frac{1}{2\kappa} \chi \frac{1}{w}(3w-1)(1+w)p =
    \frac{p}{\bar{w}}\,~, \label{rho_effective_Schutz_linear} \\
    \textrm{where} \qquad 
    &\frac{w}{\bar{w}} = 1 + \frac{1}{2\kappa} \chi (3w-1)(1+w) \,,
    \label{w2}
\end{align}
 which again shows that the linear trace of the energy-momentum contribution only changes the equation of state. Notice that in both Eqs.~\eqref{w1} and~\eqref{w2}, $\tilde{w}$ reduce to $w$ in the small $\chi$ limit, which is exactly the case of  $f(R)$ gravity with the with $p=w\rho$. On the dynamical level, we get the same field equations as $f(R)$ gravity.

\subsection{Models with linear equation of state}

Since an $f(R,T)$ model that is linear in the trace of the energy-momentum only differs from $f(R)$ gravity by the equation of state, we investigate whether there is a relation between the equations of state using the on-shell and off-shell Lagrangian for this case. Let us start by assuming a linear equation of state $p=W\rho$ in the incorrect matter term of the field equations in Eq.~\eqref{Harko_field_eqn}. This yields an effective energy density given by
\begin{align}
    \hat{\rho}_{\rm eff} := \rho + \frac{f_T}{\kappa}(\rho+p) = 
    \Bigl(1+\frac{f_T}{\kappa}(1+W)\Bigr)\rho \,,
\end{align}
where the hat in $\hat{\rho}_{\rm eff}$ is to indicate the effective energy density obtained from Eq.~\eqref{Harko_field_eqn}. We can directly compare this effective energy density with that obtained in Eq.~\eqref{rho_effective_Schutz_linear} which is also derived considering a linear equation of state $p=w\rho$. It is stated for a general $f_T$ as
\begin{align}
    \rho_{\rm eff} = 
    \Bigl(1 + \frac{f_T}{2\kappa}\frac{1}{w}(3w-1)(1+w)\Bigr)\rho\,.
\end{align}
This means that all results using a linear equation of state $W$ in the equations of~\cite{Harko:2011kv} are equivalent to those we derived in this paper using the mapping
\begin{align}
    (1+W) = \frac{1}{2w}(3w-1)(1+w)
    \label{wWeqn}
\end{align}
in the corrected equations. For example, the popular choice of dust with $W=0$ would correspond to $w=\pm 1/\sqrt{3}$, while $W=1/3$ for radiation gives $w=1/9\pm 2\sqrt{7}/9$. Interestingly, the case for dark energy with $W=-1$ corresponds to either $w=1/3$ or $w=-1$. Thus, the case of dark energy using the original equations in~\cite{Harko:2011kv} could actually correspond to radiation in the corrected equations using the Lagrangian with velocity potentials. 

Clearly, Eq.~\eqref{wWeqn} is singular in the limit $w \to 0$ which is expected as the energy density is derived from the pressure in Schutz's velocity potential formulation and therefore the equation $\rho=p/w$ becomes singular. This can be avoided by working instead in Brown's Lagrange multiplier formulation using $p_{\rm eff}$ from Eq.~\eqref{w1}. Here, when $w=0$ we get 
\begin{align}
    p_{\rm eff} = \frac{f_T}{2\kappa}\rho \,.
\end{align}
This result is equivalent to starting with $p=\tilde{w}\rho$ and setting $w=0$ in the expression for $\tilde{w}$.

This tool allows us to salvage models with linear trace of energy-momentum tensor in $f(R,T)$ gravity by realizing that such models only affect the equation of state which we can readily map to the correct equations.

\subsection{A brief note on cosmological applications}

The original $f(R,T)$ proposal~\cite{Harko:2011kv} was motivated by modifying the gravitational field equations for the late time Universe. Assuming that the pressure of the fluid is negligible at late times $p=0$, we will briefly consider the equation of state $\rho=n$ in Brown's formalism. Furthermore, we set $f(R,T)=R+\mathcal{F}(T)$ to study the effects of the trace coupling in the context of general relativity. The effective energy-momentum tensor is readily given by~(\ref{TeffBrown}) with the effective pressure simply becoming
\begin{align}
    \label{eq_p_eff2}
    p_{\rm eff} = + \frac{1}{2\kappa}f_T \rho \,,
\end{align}
where we used~(\ref{usedTdn}) to evaluate $p_{\rm eff}$. Putting this effective pressure into the acceleration equation of standard relativistic cosmology yields
\begin{align}
    \label{acc}
    \frac{\ddot{a}}{a} = -\frac{\kappa}{6}(\rho + 3p_{\rm eff}) =
    -\frac{\kappa}{6}\Bigl(\rho+\frac{3}{2\kappa}f_T \rho\Bigr) =
    -\frac{\kappa}{6}\rho\Bigl(1+\frac{3}{2\kappa}f_T\Bigr) \,.
\end{align}
This yields a neat condition on the function $f$ that corresponds to cosmological epochs of accelerated expansion, namely, 
\begin{align}
    \label{acc2}
    \frac{3}{2\kappa}f_T < -1 \,.
\end{align}
However, this is completely expected for models of this type, as suggested in Sec.~\ref{sec:summodels}. Therefore Eq.~(\ref{acc2}) is nothing but the condition $w_{\rm eff} < -1/3$ rewritten in a way that suggests that models of this form cannot contain new information. 

\subsection{Nontrivial models}

An immediate implication our work is that only models where $R$ and $T$ couple directly will contain some potentially interesting new properties for fluid models. More mathematically speaking, we require the function $f$ to satisfy $f_{TR} \neq 0$. This follows from the conservation equation~\eqref{eq_effective_divergence} which we state again in a slightly different form:
\begin{align}
    \nabla^\mu T^{\rm eff}_{\mu\nu} = 
    -\frac{1}{2} f_T(R,T) \Bigl(\frac{\partial T}{\partial n}\partial_\nu n + \frac{\partial T}{\partial s}\partial_\nu s \Bigr) \,,
\end{align}
where we emphasize that the right-hand side now does depend on the geometry through the Ricci scalar. In other words, $f_T$ must be a function of $R$.

One of the most interesting aspects of models of this type is that entropy perturbations could play a role in cosmological perturbation theory. Since $U^\nu \nabla^\mu T^{\rm eff}_{\mu\nu} = 0$ (as shown in Sec.~\ref{secconsfield}), these entropy perturbations would only affect the perpendicular part of the conversation equation. In Schutz's velocity potentials approach, we also encounter entropy perturbations. However, in this picture (as seen in Sec.~\ref{Schutz Conservation equations via field equations}), the perturbations affect both the parallel and perpendicular directions of the conservation equation.

\section{CONCLUSION}

We have demonstrated the importance of using the proper fluid Lagrangian when attempting to derive the gravitational field equations from an action principle. The use of the off-shell Lagrangian fully describes the fluid and leads to the correct equations of motion. In contrast, we showed that the use of the on-shell Lagrangian misses physical contributions that emerge from the physical variables, like entropy. Once the fluid's equations of motion are derived from the action variation, they can be subsequently used to retrieve the effective on-shell expression.

There does not exist a unique Lagrangian expression for a relativistic perfect fluid, as several expressions lead to the correct fluid equation. However, we have shown that one can reach the same conclusions regardless of the Lagrangian expression, given an appropriate off-shell Lagrangian is considered. The field equations in the two approaches are not identical, as can be seen in Eqs.~(\ref{eq_our_field_eqn}) and~(\ref{eq_Schutz_field_eqn}). However, they contain the same physical information for a given equation of state. This is in contrast to~\cite{Haghani:2023uad} where it is stated that the field equations are identical. The inequivalence of the field equations goes back to the choice of the derived quantity. When varying the energy-momentum tensor, as is required when trace couplings are considered, the variation of the derived quantity ($p$ or $\rho$) leads to second-order derivatives in the other quantity; see Eq.~\eqref{usedTdn} or~\eqref{Schutz_T_variation}. The choice of the derived quantity depends on whether one uses Brown or Schutz's approach and thus leads to the, in fact expected, difference in the field equations.    

We have shown this by considering the fluid Lagrangian given by Brown using Lagrange multipliers and also Schutz's Lagrangian expression using velocity potentials. The former is considered better as all the independent variables are explicitly expressed in the Lagrangian function and each variable is associated to a physical quantity. Yet, the expression using velocity potentials is equally valid and can have the advantage of having a simpler form to work with as the variables are hidden within the velocity function. By checking the conservation equations, we notice that the divergence of energy-momentum tensor does not vanish and indeed the coupling of geometry with the matter trace leads to contributions that do not preserve the conservation equations, most notably the entropy contributions. It has been shown that entropy perturbations in cosmological fluids are associated with large-scale perturbations in curvature hypersurfaces in multifluid interactions~\cite{Malik:2002jb,Malik:2004tf}, a topic to be further investigated in future work. The conservation equations have also been verified by deriving them both from the field equations and the fluid's equations of motion, further confirming their validity and internal consistency. Additionally, we attempted to identify the new contributions with dark matter and dark energy to better understand their possible interaction. The Schutz formulation lends itself quite naturally to yield a dark matter contribution.

The most significant implication of the findings is that $f(R,T)$ models with linear expressions of the matter trace do not lead to new physics compared to their $f(R)$ counterparts. They only lead to modified equations of state. Thus, studying the physics of nonminimal trace coupling requires the consideration of strong geometry-matter couplings or higher-order models. On a positive note, this finding means that, since models liner in the matter trace only affect the equation of state, we can map the results from such models done using the on-shell Lagrangian to the proper equations of state.

Our results apply to any modified theory of gravity that contains nonminimal couplings with perfect fluid matter. We expect that many results in this direction require a careful reinvestigation. 

\section*{ACKNOWLEDGMENTS}

We thank Tiberiu Harko and Francisco Lobo for their useful comments on this preprint. E.A. acknowledges the funding provided by Kuwait University through its graduate student scholarship program.

\section*{DATA AVAILABILITY}
No data were created or analyzed in this study.
\appendix
\section{DETAILED DERIVATIONS}

Here, we show detailed steps of some of the derivations presented in the main text. 

\subsection{Variation of the number density $n$}
\label{appen_var_n}

The variation of the number density $n$ was taken in the following steps. Starting with Eq.~\eqref{defnofJandn}, we have 
\begin{align}
    n =& \frac{\sqrt{-g_{\mu\nu} J^\mu J^\nu}}{\sqrt{-g}} \,.
\end{align}
It is more convenient to vary $n^2$, which gives
\begin{align}
    \delta n^2 =& -g_{\mu\nu} J^\mu J^\nu \delta\Big(\frac{1}{-g}\Big) - \frac{J^\mu J^\nu}{-g} \delta \big(g_{\mu\nu}\big)-2\frac{g_{\mu\nu}}{-g}J^\nu \delta J^\mu \,, \notag \\
    2n\delta n =& -g_{\mu\nu} J^\mu J^\nu \Big(\frac{-gg_{\alpha\beta}\delta g^{\alpha\beta}}{g^2}\Big) - \frac{J^\mu J^\nu}{-g} \Big(-\frac{1}{2}\big(g_{\mu\alpha}g_{\nu\beta}+g_{\mu\beta}g_{\nu\alpha}\big)\delta g^{\alpha \beta}\Big)-2\frac{g_{\mu\nu}}{-g}J^\nu \delta J^\mu \notag \\
    =& - J_\nu J^\nu \frac{g_{\alpha\beta}\delta g^{\alpha\beta}}{-g} + \frac{J_\alpha J_\beta}{-g}\delta g^{\alpha \beta}-2\frac{J_\mu}{-g}\delta J^\mu \notag \\
    =& - \big(\sqrt{-g}nU_\nu\big)\big(\sqrt{-g}nU^\nu\big) \frac{g_{\alpha\beta}\delta g^{\alpha\beta}}{-g} + \frac{\big(\sqrt{-g}nU_\alpha\big)\big(\sqrt{-g}nU_\beta\big)}{-g}\delta g^{\alpha \beta}-2\frac{\big(\sqrt{-g}nU_\mu\big)}{-g}\delta J^\mu \notag \\
    =&  n^2g_{\alpha\beta}\delta g^{\alpha\beta} + n^2 U_\alpha U_\beta\delta g^{\alpha \beta}-2\frac{nU_\mu}{\sqrt{-g}}\delta J^\mu \,, \notag \\
    \Rightarrow \quad \delta n =& \frac{n}{2} \big(g_{\alpha\beta} + U_\alpha U_\beta \big)\delta g^{\alpha\beta} - \frac{U_\mu}{\sqrt{-g}}\delta J^\mu \label{eq_delta_n} \,.
\end{align}
We use $\delta n$ when we derive the variation of the trace of the energy-momentum tensor $\delta T$ in Brown's approach.

\subsection{Variation of Brown's fluid Lagrangian}
\label{appen_var_L}

Starting from the variation of the fluid Lagrangian, Eq.~\eqref{eq_Lagrangian_var_1}, we have
\begin{align}
    \delta \mathcal{L}_{\mathrm{fluid}} =& -\rho\delta \sqrt{-g} - \sqrt{-g}\delta \rho + (\varphi_{,\mu}+s\theta_{,\mu}+\beta_A\alpha^A_{,\mu})\delta J^\mu + J^\mu \delta(\varphi_{,\mu}+s\theta_{,\mu}+\beta_A\alpha^A_{,\mu}) \notag \\
    =& -\rho \big(-\frac{1}{2}\sqrt{-g}g_{\mu\nu}\delta g^{\mu\nu} \big) - \sqrt{-g}\big( \frac{\partial \rho}{\partial n}\delta n + \frac{\partial \rho}{\partial s}\delta s\big) + (\varphi_{,\mu}+s\theta_{,\mu}+\beta_A\alpha^A_{,\mu})\delta J^\mu \notag \\&+ J^\mu \delta(\varphi_{,\mu}+s\theta_{,\mu}+\beta_A\alpha^A_{,\mu}) \,.
\end{align}
Substituting the expression for $\delta n$ derived in Eq.~\eqref{eq_delta_n},
\begin{align}
   \delta \mathcal{L}_{\mathrm{fluid}} =& \frac{1}{2}\rho \sqrt{-g} g_{\mu\nu}\delta g^{\mu\nu} - \sqrt{-g}\frac{\partial \rho}{\partial n}\Big(\frac{n}{2} \big(g_{\mu\nu} + U_\mu U_\nu \big)\delta g^{\mu\nu} - \frac{U_\mu}{\sqrt{-g}}\delta J^\mu \Big) - \sqrt{-g} \frac{\partial \rho}{\partial s}\delta s \notag \\
    &+ (\varphi_{,\mu}+s\theta_{,\mu}+\beta_A\alpha^A_{,\mu})\delta J^\mu +J^\mu \delta\varphi_{,\mu}+J^\mu s\delta\theta_{,\mu}+ J^\mu\theta_{,\mu}\delta s+J^\mu\beta_A\delta\alpha^A_{,\mu}+J^\mu\alpha^A_{,\mu}\delta \beta_A) \notag \\
    =& \frac{1}{2}\sqrt{-g}\Big(\rho g_{\mu\nu} - n \frac{\partial \rho}{\partial n}\big(g_{\mu\nu} + U_\mu U_\nu \big) \Big)\delta g^{\mu\nu}+\Big( \frac{\partial \rho}{\partial n}U_\mu + (\varphi_{,\mu}+s\theta_{,\mu}+\beta_A\alpha^A_{,\mu})\Big)\delta J^\mu \notag \\
    &+\Big(J^\mu \theta_{,\mu} - \sqrt{-g}\frac{\partial \rho}{\partial s}\Big) \delta s+J^\mu \delta\varphi_{,\mu}+J^\mu s\delta\theta_{,\mu}+J^\mu\beta_A\delta\alpha^A_{,\mu}+J^\mu\alpha^A_{,\mu}\delta \beta_A \,.
\end{align}
Using integration by parts, we arrive at
\begin{align}
   \delta \mathcal{L}_{\mathrm{fluid}} =& \frac{1}{2}\sqrt{-g}\Big(\rho g_{\mu\nu} - n \frac{\partial \rho}{\partial n}\big(g_{\mu\nu} + U_\mu U_\nu \big) \Big)\delta g^{\mu\nu}+\Big( \frac{\partial \rho}{\partial n}U_\mu + (\varphi_{,\mu}+s\theta_{,\mu}+\beta_A\alpha^A_{,\mu})\Big)\delta J^\mu \notag \\
    &+\sqrt{-g}\Big(nU^\mu \theta_{,\mu} -\frac{\partial \rho}{\partial s}\Big) \delta s - J^\mu_{,\mu} \delta\varphi-\big(J^\mu s\big)_{,\mu}\delta\theta-\big(J^\mu\beta_A\big)_{,\mu}\delta\alpha^A+J^\mu\alpha^A_{,\mu}\delta \beta_A \,,
\end{align}
which is the expression provided in Eq.~\eqref{eq_L_variation}.

\subsection{Conservation equations via field equations in Brown's approach}
\label{appen_cons_field}

We show here the detailed derivation of Eq.~\eqref{hconservation1a}. Starting from Eq.~\eqref{eq_perpen_cons_field}, we rewrite it as
\begin{align}
     2\kappa h^{\nu\alpha} \nabla^\mu T_{\mu\nu} &=  h^{\alpha}_\mu{} \nabla^\mu(f_T n\frac{\partial T}{\partial n}) - f_T h^{\nu\alpha} \nabla_\nu T + f_T n\frac{\partial T}{\partial n} h^{\nu\alpha}\nabla^\mu h_{\mu\nu} \,.
\end{align}
Next, use $h^{\nu\alpha}\nabla^\mu h_{\mu\nu} = U^\mu \nabla_\mu U^\alpha$:
\begin{align}
     2\kappa h^{\nu\alpha} \nabla^\mu T_{\mu\nu} &=   h^{\alpha}_\mu{} \nabla^\mu(f_T n\frac{\partial T}{\partial n}) -  f_T h^{\nu\alpha} \nabla_\nu T + f_T n\frac{\partial T}{\partial n} U^\mu \nabla_\mu U^\alpha \,,
\end{align}
Therefore
\begin{align}
     2\kappa h^{\nu\alpha} \nabla^\mu T_{\mu\nu} =  
       n& h^{\alpha}_\mu{}
     \nabla^\mu \Big(f_T\frac{\partial T}{\partial n}\Big)+f_T \frac{\partial T}{\partial n} h^{\alpha}_\mu{}\nabla^\mu n -
     f_T h^{\nu\alpha} \Big(\frac{\partial T}{\partial n}\nabla_\nu n +
     \frac{\partial T}{\partial s} \nabla_\nu s \Big)
      +  f_T n\frac{\partial T}{\partial n} U^\mu \nabla_\mu U^\alpha \notag \\
     =
     n&
     h^{\alpha}_\mu{}\nabla^\mu (f_T \frac{\partial T}{\partial n}) -
     f_T h^{\nu\alpha} \frac{\partial T}{\partial s} \nabla_\nu s
      +  f_T n\frac{\partial T}{\partial n} U^\mu \nabla_\mu U^\alpha \,.
\end{align}
Rename and lower one index:
\begin{align}
    2\kappa h^{\nu}_{\alpha} \nabla^\mu T_{\mu\nu} &=
       n
     h_{\alpha}^{\mu}\nabla_\mu (f_T \frac{\partial T}{\partial n}) -
     f_T \frac{\partial T}{\partial s} \nabla_\alpha s
      +  f_T n\frac{\partial T}{\partial n} U^\mu \nabla_\mu U_\alpha \,.
\end{align}
This is Eq.~\eqref{hconservation1a}, given in the main text.

\subsection{Conservation equations via fluid equations in Brown's approach}
\label{appen_cons_fluid}

We approach the perpendicular contraction by starting from the divergence of the energy-momentum tensor. First, we show the derivation leading to Eq.~\eqref{hconservation2}, which is also given in~\cite{Brown_1993}. The divergence of the energy-momentum tensor is
\begin{align}
   \nabla_\beta T^{\alpha \beta} &= \nabla_\beta\big(\rho U^\alpha U^\beta\big)+\nabla_\beta\Big(p\big(g^{\alpha \beta}+U^\alpha U^\beta\big)\Big) \,.
\end{align}
Now apply $h_{\alpha\sigma}=g_{\alpha\sigma}+U_\alpha U_\sigma$:
\begin{align}
   h_{\alpha \sigma}\nabla_\beta T^{\alpha \beta}=h_{\alpha \sigma}\bigg[&\nabla_\beta\big(\rho U^\alpha U^\beta\big)+\nabla_\beta\Big(p\big(g^{\alpha \beta}+U^\alpha U^\beta\big)\Big)\bigg] \notag \\
   = h_{\alpha \sigma}\bigg[&U^{\alpha}U^\beta \nabla_\beta \rho + \rho U^\alpha \nabla_\beta U^\beta+\rho U^\beta\nabla_\beta U^\alpha +(g^{\alpha\beta}+U^\alpha U^\beta)\nabla_\beta p \notag \\
     &+ pU^\alpha\nabla_\beta U^\beta + p U^\beta \nabla_\beta U^\alpha\bigg] \,.
\end{align}
Since $h_{\alpha\sigma}U^\alpha=0$, the first, second, and fifth terms in the square brackets vanish. We can write
\begin{align}
    h_{\alpha \sigma}\nabla_\beta T^{\alpha \beta} &= \big(U_\alpha U_\sigma +g_{\alpha\sigma}\big)\bigg[\rho U^\beta\nabla_\beta U^\alpha + (g^{\alpha\beta}+U^\alpha U^\beta)\nabla_\beta p+p U^\beta \nabla_\beta U^\alpha\bigg] \,~.
\end{align}
Now, apply the contraction and combine terms:
\begin{align}
   h_{\alpha \sigma}\nabla_\beta T^{\alpha \beta} &= \big(U_\alpha U_\sigma +g_{\alpha\sigma}\big)\rho U^\beta\nabla_\beta U^\alpha + (\delta^{\beta}_{\sigma}+U_\sigma U^\beta)\nabla_\beta p+\big(U_\alpha U_\sigma +g_{\alpha\sigma}\big)p U^\beta \nabla_\beta U^\alpha
   \notag \\
   &= \big(\rho+p\big)\big(U_\alpha U_\sigma +g_{\alpha\sigma}\big) U^\beta\nabla_\beta U^\alpha +(\delta^{\beta}_{\sigma}+U_\sigma U^\beta)\nabla_\beta p \notag \\
   &= \big(\rho+p\big)U^\beta \nabla_\beta U_{\sigma} +(\delta^{\beta}_{\sigma}+U_\sigma U^\beta)\nabla_\beta p \,,
\end{align}
where we have $U_\alpha \nabla_\beta U^\alpha=0$. By substituting $\upmu=(\rho+p)/n$, we  get
\begin{align}
    h_{\alpha \sigma}\nabla_\beta T^{\alpha \beta} &= n\upmu U^\beta\nabla_\beta U_\sigma +(\delta^{\beta}_{\sigma}+U_\sigma U^\beta)\nabla_\beta p
   \notag \\
   &= n\Big[U^\beta\nabla_\beta (\upmu U_\sigma)-U^\beta U_\sigma\nabla_\beta\upmu \Big]  +(\delta^{\beta}_{\sigma}+U_\sigma U^\beta)\nabla_\beta p \,.
\end{align}
We now add terms $\pm U^\beta\nabla_\sigma (\upmu U_\beta)$ which allows us to antisymmetrize the first term:
\begin{align}
   h_{\alpha \sigma}\nabla_\beta T^{\alpha \beta} &= n\Big[U^\beta\nabla_\beta (\upmu U_\sigma)-U^\beta\nabla_\sigma (\upmu U_\beta)+U^\beta\nabla_\sigma (\upmu U_\beta)-U^\beta U_\sigma\nabla_\beta\upmu \Big] +(\delta^{\beta}_{\sigma}+U_\sigma U^\beta)\nabla_\beta p
   \notag \\
   &= n\Big[2U^\beta\nabla_{[\beta} (\upmu U_{\sigma]})+\upmu U^\beta\nabla_\sigma U_\beta+U^\beta U_\beta\nabla_\sigma\upmu -U^\beta U_\sigma\nabla_\beta\upmu  +(\delta^{\beta}_{\sigma}+U_\sigma U^\beta)\frac{1}{n}\nabla_\beta p\Big] \notag \\
   &= n\Big[2U^\beta\nabla_{[\beta} (\upmu U_{\sigma]})-\nabla_\sigma\upmu -U^\beta U_\sigma\nabla_\beta\upmu  +(\delta^{\beta}_{\sigma}+U_\sigma U^\beta)\frac{1}{n}\nabla_\beta p\Big]
   \notag \\
   &= n\Big[2U^\beta\nabla_{[\beta} (\upmu U_{\sigma]})-(\delta^{\beta}_{\sigma}+U_\sigma U^\beta)\nabla_\beta\upmu +(\delta^{\beta}_{\sigma}+U_\sigma U^\beta)\frac{1}{n}\nabla_\beta p\Big]
   \notag \\
   &= n\Big[2U^\beta\nabla_{[\beta} (\upmu U_{\sigma]})-(\delta^{\beta}_{\sigma}+U_\sigma U^\beta)\big(\nabla_\beta\upmu-\frac{1}{n}\nabla_\beta p\big)\Big]
   \notag \\
   &= n\Big[2U^\beta\nabla_{[\beta} (\upmu U_{\sigma]})-(\delta^{\beta}_{\sigma}+U_\sigma U^\beta)\big(\frac{1}{n}\nabla_\beta(\rho+p)-\frac{1}{n^2}(\rho + p)\nabla_\beta n-\frac{1}{n}\nabla_\beta p\big)\Big]
   \notag \\
   &= n\Big[2U^\beta\nabla_{[\beta} (\upmu U_{\sigma]})-(\delta^{\beta}_{\sigma}+U_\sigma U^\beta)\big(\frac{1}{n}\frac{\partial \rho}{\partial n}\nabla_\beta n+\frac{1}{n}\frac{\partial \rho}{\partial s}\nabla_\beta s-\frac{1}{n^2}(\rho + p)\nabla_\beta n\big)\Big]
   \notag \\
   &= n\Big[2U^\beta\nabla_{[\beta} (\upmu U_{\sigma]})-(\delta^{\beta}_{\sigma}+U_\sigma U^\beta)\frac{1}{n}\frac{\partial \rho}{\partial s}\nabla_\beta s\Big]
   \notag \\
   &= 2nU^\beta\nabla_{[\beta} (\upmu U_{\sigma]})-h_\sigma^\beta\frac{\partial \rho}{\partial s}\nabla_\beta s \,.
\end{align}
We can now rename, lower, and raise indices on both sides to get 
\begin{align}
\label{eq_appen_perpen_div}
     h^{\nu}_{\alpha} \nabla^\mu T_{\mu\nu} =
     2nU^\beta\nabla_{[\beta}(\upmu U_{\alpha]}) - 
      \frac{\partial\rho}{\partial s}\nabla_\alpha s \,.
\end{align}
We can find an expression for the first term on the right-hand side. First, consider the relation in Eq.~\eqref{Jeq}:
\begin{align}
    \upmu U_\alpha - \frac{f_T}{2\kappa}\Bigl(3 n\frac{\partial^2 \rho}{\partial n^2}
    - \frac{\partial \rho}{\partial n}\Bigr) U_\alpha + 
    \varphi_{,\alpha}+s\theta_{,\alpha}+\beta_A\alpha^A_{,\alpha} = 0 \,.
\end{align}
Taking the skew part of the divergence of this equation,
\begin{align}
    \nabla_{[\beta}(\upmu U_{\alpha]}) - \nabla_{[\beta}\Bigg(\frac{f_T}{2\kappa}\Bigl(3 n\frac{\partial^2 \rho}{\partial n^2}
    - \frac{\partial \rho}{\partial n}\Bigr) U_{\alpha]}\Bigg) + \nabla_{[\beta}\varphi_{,\alpha]}+\nabla_{[\beta}(s\theta_{,\alpha]})+\nabla_{[\beta}(\beta_A\alpha^A_{,\alpha]}) = 0 \,.
\end{align}
We multiply by $2U^\beta$:
\begin{multline}
\label{eq_appen_note1}
     2U^\beta\nabla_{[\beta}(\upmu U_{\alpha]}) -2U^\beta\nabla_{[\beta}\bigg(\frac{f_T}{2\kappa}\Bigl(3 n\frac{\partial^2 \rho}{\partial n^2}
    - \frac{\partial \rho}{\partial n}\Bigr) U_{\alpha]}\bigg)\\ + 2U^\beta\nabla_{[\beta}\varphi_{,\alpha]}+2U^\beta\nabla_{[\beta}(s\theta_{,\alpha]})+2U^\beta\nabla_{[\beta}(\beta_A\alpha^A_{,\alpha]}) = 0 \,.
\end{multline}
We note
\begin{align}
    2U^\beta\nabla_{[\beta}\varphi_{,\alpha]} &= 
    U^\beta(\nabla_{\beta}\nabla_{\alpha}\varphi-\nabla_{\alpha}\nabla_{\beta}\varphi) = 0 \,, \\
    2U^\beta\nabla_{[\beta}(s\theta_{,\alpha]}) &= 
    U^\beta\Big(\nabla_{\beta}(s\theta_{,\alpha}) -
    \nabla_{\alpha}(s\theta_{,\beta})\Big) \notag \\ &=
    U^\beta\Big(\nabla_{\beta} s \nabla_\alpha \theta + 
     s \nabla_{\alpha}\nabla_\beta \theta -
    \nabla_{\alpha} s \nabla_\beta \theta -
     s \nabla_{\alpha} \nabla_\beta \theta \Big) \notag \\ &=
    U^\beta\Big(\nabla_{\beta} s \nabla_\alpha \theta -
    \nabla_{\alpha} s \nabla_\beta \theta\Big) \notag \\
    &=-U^\beta\nabla_{\alpha} s \nabla_\beta \theta \,, \qquad \qquad\text{using Eq.~\eqref{thetaeq}}\,, \notag \\
    2U^\beta\nabla_{[\beta}(\beta_A\alpha^A_{,\alpha]})&= U^\beta\Big(\nabla_{\beta}(\beta_A\nabla_\alpha \alpha^A) -\nabla_{\alpha}(\beta_A \nabla_\beta\alpha^A)\Big) \notag \\
     &= U^\beta \Big(\nabla_{\beta}\beta_A \nabla_\alpha \alpha^A + 
    \beta_A \nabla_{\beta} \nabla_\alpha \alpha^A -
    \nabla_{\alpha}\beta_A \nabla_\beta\alpha^A -
    \beta_A \nabla_{\alpha} \nabla_\beta\alpha^A \Big) \notag \\ &= 
    U^\beta\Big(\nabla_{\beta}\beta_A \nabla_\alpha \alpha^A -
    \nabla_{\alpha}\beta_A \nabla_\beta\alpha^A \Big) \notag \\
     &= U^\beta \nabla_{\beta}\beta_A \nabla_\alpha \alpha^A -
    U^\beta \nabla_{\alpha}\beta_A \nabla_\beta\alpha^A \notag \\
    &= 0 \,, \qquad  \qquad \text{using Eqs.~\eqref{alphaeq} and~\eqref{betaeq}}\,.
\end{align}
Thus, Eq.~\eqref{eq_appen_note1} becomes
\begin{align}
     2U^\beta\nabla_{[\beta}(\upmu U_{\alpha]}) -2U^\beta\nabla_{[\beta}\bigg(\frac{f_T}{2\kappa}\Bigl(3 n\frac{\partial^2 \rho}{\partial n^2}
    - \frac{\partial \rho}{\partial n}\Bigr) U_{\alpha]}\bigg)-U^\beta\nabla_{\beta}\theta \nabla_\alpha s = 0 \,.
\end{align}
Multiplying by $n$ and using Eq.~\eqref{seq},
\begin{align}
    2nU^\beta\nabla_{[\beta}(\upmu U_{\alpha]}) -2nU^\beta\nabla_{[\beta}\bigg(\frac{f_T}{2\kappa}\Bigl(3 n\frac{\partial^2 \rho}{\partial n^2}
    - \frac{\partial \rho}{\partial n}\Bigr) U_{\alpha]}\bigg)-  \Big(\frac{\partial \rho}{\partial s}-\frac{1}{2\kappa}f_T\frac{\partial T}{\partial s}\Big)\nabla_\alpha s = 0 \,.
\end{align}
Thus,
\begin{align}
    2nU^\beta\nabla_{[\beta}(\upmu U_{\alpha]}) = 2nU^\beta\nabla_{[\beta}\bigg(\frac{f_T}{2\kappa}\Bigl(3 n\frac{\partial^2 \rho}{\partial n^2}
    - \frac{\partial \rho}{\partial n}\Bigr) U_{\alpha]}\bigg)+  \Big(\frac{\partial \rho}{\partial s}-\frac{1}{2\kappa}f_T\frac{\partial T}{\partial s}\Big)\nabla_\alpha s \,.
\end{align}
Now substituting back in Eq.~\eqref{eq_appen_perpen_div}gives
\begin{align}
    h^{\nu}_{\alpha} \nabla^\mu T_{\mu\nu} &=
    2nU^\beta\nabla_{[\beta}\bigg(\frac{f_T}{2\kappa}\Bigl(3 n\frac{\partial^2 \rho}{\partial n^2}
    - \frac{\partial \rho}{\partial n}\Bigr) U_{\alpha]}\bigg)+  \Big(\frac{\partial \rho}{\partial s}-\frac{1}{2\kappa}f_T\frac{\partial T}{\partial s}\Big)\nabla_\alpha s - \frac{\partial \rho}{\partial s}\nabla_\alpha s
    \notag \\
    &= 2nU^\beta \nabla_{[\beta}\Bigg(\frac{f_T}{2\kappa} \Bigl(3 n\frac{\partial^2 \rho}{\partial n^2} - \frac{\partial \rho}{\partial n}\Bigr) U_{\alpha]}\Bigg) -
    \frac{f_T}{2\kappa}\frac{\partial T}{\partial s}\nabla_\alpha s \notag \\
    &= 2nU^\beta \nabla_{[\beta}\Big(\frac{f_T}{2\kappa}\frac{\partial T}{\partial n} U_{\alpha]}\Big) -
    \frac{f_T}{2\kappa}\frac{\partial T}{\partial s}\nabla_\alpha s \notag \\
    &= nU^\beta \nabla_{\beta}\Big(\frac{f_T}{2\kappa}\frac{\partial T}{\partial n} U_{\alpha}\Big)-nU^\beta \nabla_{\alpha}\Big(\frac{f_T}{2\kappa}\frac{\partial T}{\partial n} U_{\beta}\Big) -
    \frac{f_T}{2\kappa}\frac{\partial T}{\partial s}\nabla_\alpha s \notag \\
    &= nU^\beta U_\alpha \nabla_{\beta}\Big(\frac{f_T}{2\kappa}\frac{\partial T}{\partial n} \Big)+n\frac{f_T}{2\kappa}\frac{\partial T}{\partial n}U^\beta \nabla_{\beta}U_{\alpha}-nU^\beta U_{\beta} \nabla_{\alpha}\Big(\frac{f_T}{2\kappa}\frac{\partial T}{\partial n} \Big)\notag \\
    &-n\frac{f_T}{2\kappa}\frac{\partial T}{\partial n} U^\beta \nabla_{\alpha} U_{\beta} -
    \frac{f_T}{2\kappa}\frac{\partial T}{\partial s}\nabla_\alpha s
\end{align}
where, since $U^\beta \nabla_\alpha U_\beta=0$, we get
\begin{align}
   h^{\nu}_{\alpha} \nabla^\mu T_{\mu\nu} &= nU^\beta U_\alpha \nabla_{\beta}\Big(\frac{f_T}{2\kappa}\frac{\partial T}{\partial n} \Big)+n\frac{f_T}{2\kappa}\frac{\partial T}{\partial n}U^\beta \nabla_{\beta}U_{\alpha} + n \delta^\beta_\alpha\nabla_{\beta}\Big(\frac{f_T}{2\kappa}\frac{\partial T}{\partial n} \Big) - \frac{f_T}{2\kappa}\frac{\partial T}{\partial s}\nabla_\alpha s
    \notag \\
    &= n h^\beta_\alpha\nabla_{\beta}\Big(\frac{f_T}{2\kappa}\frac{\partial T}{\partial n} \Big) +n\frac{f_T}{2\kappa}\frac{\partial T}{\partial n}U^\beta \nabla_{\beta}U_{\alpha} - \frac{f_T}{2\kappa}\frac{\partial T}{\partial s}\nabla_\alpha s \,.
\end{align}
Thus,
\begin{align}
    2\kappa h^{\nu}_{\alpha} \nabla^\mu T_{\mu\nu} &=
    h^\beta_{\alpha} n \nabla_{\beta}\Big(f_T \frac{\partial T}{\partial n} \Big) + n f_T \frac{\partial T}{\partial n} U^\beta \nabla_{\beta} U_{\alpha} -
    f_T\frac{\partial T}{\partial s}\nabla_\alpha s \,,
\end{align}
which is Eq.~\eqref{eq_perpen_fluid_cons}. 

One can rewrite this using $h^{\beta}_{\alpha}\nabla_\mu h^\mu_{\beta}=U^\beta \nabla_\beta U_\alpha$ to get
\begin{align}
    2\kappa h^{\nu}_{\alpha} \nabla^\mu T_{\mu\nu} &= n
    h^\beta_{\sigma}h^\sigma_{\alpha} \nabla_{\beta}\Big(f_T \frac{\partial T}{\partial n} \Big) + n f_T \frac{\partial T}{\partial n} h^\sigma_{\alpha} \nabla_{\beta} h^{\beta}_{\sigma} -
    f_T\frac{\partial T}{\partial s}\nabla_\alpha s
    \notag \\
    &= n h^\sigma_{\alpha} \Bigg(  h^\beta_{\sigma} \nabla_{\beta}\Big(f_T \frac{\partial T}{\partial n} \Big) + n f_T \frac{\partial T}{\partial n} \nabla_{\beta} h^{\beta}_{\sigma} \Bigg)-
    f_T\frac{\partial T}{\partial s}\nabla_\alpha s
    \notag \\
    &= n h^\sigma_{\alpha} \nabla_{\beta}\Big(h^\beta_{\sigma}f_T \frac{\partial T}{\partial n} \Big)  -
    f_T\frac{\partial T}{\partial s}\nabla_\alpha s
    \notag \\ 
    &= h^\sigma_{\alpha} \Big(n \nabla_{\beta}\Big(h^\beta_{\sigma}f_T \frac{\partial T}{\partial n} \Big)  -
    f_T\frac{\partial T}{\partial s}\nabla_\sigma s \Big) \,,
\end{align}
which is a much nicer way to present the term for the divergence perpendicular to the flow.
 
\subsection{Conservation equations via fluid equations in Schutz's approach}
\label{appendix_vanishing_relation}

We show how the conservation relation for the effective energy-momentum tensor in Eq.~\eqref{eq_effective_divergence} can be verified using the fluid equation. While the parallel contraction is trivial, the perpendicular is not and we present a detailed account of it here.

Now, for the perpendicular contraction, we want to show that 
\begin{align}
h^\nu_\sigma\nabla^\mu T_{\mu\nu}^{\rm eff}=-\frac{1}{2\kappa}f_Th^\nu_\sigma\nabla_\nu T \,.
\end{align}
Starting from the effective fluid equation,
\begin{align}
    \nabla^\mu T_{\mu\nu}^{\rm eff} &= \nabla^\mu \Big(\upmu nU_\mu U_\nu+p g_{\mu\nu}\Big) \notag \\
    &= \upmu nU_\mu\nabla^\mu U_\nu + n U_\mu U_\nu \nabla^\mu \upmu +g_{\mu\nu} \nabla^\mu p \,, \\
    h^\nu_\sigma\nabla^\mu T_{\mu\nu}^{\rm eff} &= \upmu n U_\mu h^\nu_\sigma \nabla^\mu U_\nu + h^\nu_\sigma \nabla_\nu p \notag \\
    &= \upmu n U_\mu h^\nu_\sigma \nabla^\mu U_\nu +  h^\nu_\sigma\frac{\partial p}{\partial \upmu} \nabla_\nu \upmu +  h^\nu_\sigma\frac{\partial p}{\partial s} \nabla_\nu s \notag \\
    &= \upmu n U_\mu h^\nu_\sigma \nabla^\mu U_\nu + n h^\nu_\sigma\nabla_\nu \upmu-\frac{f_T}{2k}\frac{\partial T}{\partial \upmu}  h^\nu_\sigma\nabla_\nu \upmu -nU^\lambda \theta_{,\lambda}  h^\nu_\sigma \nabla_\nu s - \frac{f_T}{2\kappa}\frac{\partial T}{\partial s} h^\nu_\sigma \nabla_\nu s \notag \\
    &= \upmu n U_\mu h^\nu_\sigma \nabla^\mu U_\nu + n h^\nu_\sigma\nabla_\nu \upmu -nU^\lambda \theta_{,\lambda}  h^\nu_\sigma \nabla_\nu s - \frac{f_T}{2\kappa} h^\nu_\sigma\nabla_\nu T  \,.\label{perpen_inter_eq1_appen}
\end{align}
We find that the last term is exactly what we seek and thus we need to show that 
\begin{align}
    \upmu n U_\mu h^\nu_\sigma \nabla^\mu U_\nu + n h^\nu_\sigma\nabla_\nu \upmu -nU^\lambda \theta_{,\lambda}  h^\nu_\sigma \nabla_\nu s = 0 \,.
\end{align}
If we evaluate $\nabla^\mu U_\nu$, we find
\begin{align}
    \nabla^\mu U_\nu &= \nabla^\mu \Big(\upmu^{-1}\big(\partial_\nu\phi+\alpha\partial_\nu\beta+\theta\partial_\nu s\big)\Big) \notag \\
    &= -\upmu^{-2}\big(\partial_\nu\phi+\alpha\partial_\nu\beta+\theta\partial_\nu s \big) \nabla^\mu \upmu \notag \\& \qquad + \upmu^{-1}\big(\partial^\mu\partial_\nu\phi+\partial^\mu\alpha\partial_\nu\beta+\alpha\partial^\mu\partial_\nu\beta+\partial^\mu\theta\partial_\nu s+\theta\partial^\mu\partial_\nu s \big) \notag \\
    &= -\upmu^{-1}U_\nu\nabla^\mu \upmu + \upmu^{-1}\big(\partial^\mu\partial_\nu\phi+\partial^\mu\alpha\partial_\nu\beta+\alpha\partial^\mu\partial_\nu\beta+\partial^\mu\theta\partial_\nu s+\theta\partial^\mu\partial_\nu s \big) \,,
\end{align}
and this means we have
\begin{align}
    \upmu U_\mu\nabla^\mu U_\nu = -U_\mu U_\nu \nabla^\mu \upmu + U_\mu\partial^\mu\partial_\nu\phi+U_\mu\partial^\mu\alpha\partial_\nu\beta&+U_\mu\alpha\partial^\mu\partial_\nu\beta \notag \\&+U_\mu\partial^\mu\theta\partial_\nu s+U_\mu\theta\partial^\mu\partial_\nu s \notag \\ 
    = -U_\mu U_\nu \nabla^\mu \upmu + U_\mu\partial^\mu\partial_\nu\phi+U_\mu\alpha\partial^\mu\partial_\nu\beta&+U_\mu\partial^\mu\theta\partial_\nu s+U_\mu\theta\partial^\mu\partial_\nu s \label{appen_exp1} \,,
\end{align}
where we used Eq.~\eqref{Schutz_delta_beta} to write $U_\mu \partial^\mu \alpha=0$. Now, using the equations of motion and the product rule, we find that
\begin{align}
\label{appen_alpha_relation}
    \nabla_\nu(U_\mu \nabla^\mu \alpha) = 0 \Rightarrow U_\mu  \nabla_\nu\nabla^\mu \alpha =- \nabla^\mu\alpha\nabla_\nu U_\mu \,.
\end{align}
Similarly, one can show
\begin{align}
    U_\mu\nabla_\nu \nabla^\mu  \beta &=- \nabla^\mu\beta\nabla_\nu U_\mu \,,\label{appen_beta_relation} \\
     U_\mu\nabla_\nu \nabla^\mu  s &=- \nabla^\mu s \nabla_\nu U_\mu \,.\label{appen_s_relation} 
\end{align}
We also have
\begin{align}
    U_\mu \nabla^\mu \phi&=-\upmu \,,\\
    \Rightarrow \nabla_\nu( U_\mu \nabla^\mu \phi ) &= -\nabla_\nu\upmu \,, \\
    U_\mu \nabla_\nu\nabla^\mu \phi &= -\nabla_\nu\upmu - \nabla^\mu \phi\nabla_\nu U_\mu \label{appen_phi_relation} \,.
\end{align}
Now, we can use Eqs.~\eqref{appen_phi_relation},~\eqref{appen_alpha_relation}, and.~\eqref{appen_s_relation} to write Eq.~\eqref{appen_exp1} as
\begin{align}
    \upmu U_\mu\nabla^\mu U_\nu = -U_\mu U_\nu \nabla^\mu \upmu &- \nabla_\nu \upmu- \nabla_\nu U_\mu\nabla^\mu\phi-\nabla_\nu U_\mu\big(\alpha\nabla^\mu\beta) \notag \\&+U_\mu\nabla^\mu\theta\nabla_\nu s-\nabla_\nu U_\mu(\theta\nabla^\mu s) \notag \\
    = -U_\mu U_\nu \nabla^\mu \upmu &- \nabla_\nu \upmu + U_\mu\nabla^\mu\theta\nabla_\nu s - V^\mu\nabla_\nu U_\mu \notag \\
    = -U_\mu U_\nu \nabla^\mu \upmu &- \nabla_\nu \upmu + U_\mu\nabla^\mu\theta\nabla_\nu s - \upmu U^\mu\nabla_\nu U_\mu \notag \\
    = -U_\mu U_\nu \nabla^\mu \upmu &- \nabla_\nu \upmu + U_\mu\nabla^\mu\theta\nabla_\nu s \,.  \label{Schutz_velocity_relation_appen}
\end{align}
We can now substitute this expression in Eq.~\eqref{perpen_inter_eq1_appen}. We get
\begin{align}
     h^\nu_\sigma\nabla^\mu T_{\mu\nu}^{\rm eff} =  n h^\nu_\sigma \Big(-U_\mu U_\nu &\nabla^\mu \upmu - \nabla_\nu \upmu + U_\mu\nabla^\mu\theta\nabla_\nu s \Big) \notag \\&+ n h^\nu_\sigma\nabla_\nu \upmu -nU^\lambda \theta_{,\lambda}  h^\nu_\sigma \nabla_\nu s - \frac{f_T}{2\kappa} h^\nu_\sigma\nabla_\nu T 
    \notag \\ h^\nu_\sigma\nabla^\mu T_{\mu\nu}^{\rm eff}=  - \frac{f_T}{2\kappa} h^\nu_\sigma\nabla_\nu T \,,&
\end{align}
where $h^\nu_\sigma U_\nu = 0$. We indeed get the result expected.

In general, we notice that we have a relation
\begin{align}
     h^\nu_\sigma \big(\upmu U_\mu\nabla^\mu U_\nu +\nabla_\nu \upmu - U_\mu \nabla^\mu\theta\nabla_\nu s\big) = 0 \,, \label{special_appen_relation}
\end{align}
which is useful when deriving the conservation equations. This term can also be written as
\begin{align}
    h^\nu_\sigma \Big(\upmu U_\mu\nabla^\mu U_\nu +\nabla_\nu \upmu - U_\mu \nabla^\mu\theta\nabla_\nu s\Big)&= h^\nu_\sigma \Big(\upmu \nabla^\mu\big(U_\mu U_\nu \big) +g_{\mu\nu}\nabla^\mu \upmu - U_\mu \nabla^\mu\theta\nabla_\nu s \Big) \notag \\
    &= h^\nu_\sigma \Big(\upmu \nabla^\mu\big(U_\mu U_\nu +g_{\mu\nu}\big) +\big(g_{\mu\nu}+U_\mu U_\nu \big)\nabla^\mu \upmu \notag \\ & \qquad - U_\mu \nabla^\mu\theta\nabla_\nu s \Big) \notag \\
    &=h^\nu_\sigma \Big(\upmu \nabla^\mu h_{\mu\nu}+h_{\mu\nu}\nabla^\mu \upmu -U^\lambda\theta_{,\lambda}\nabla_\nu s\Big) \label{special_appen_relation_2}\,,
\end{align}
where we used the fact that the covariant derivative of the metric vanishes and $h^\nu_\sigma U_\nu = 0$. 

\providecommand{\href}[2]{#2}\begingroup\raggedright\endgroup

%\addcontentsline{toc}{section}{References}
%\bibliographystyle{jhepmodstyle}
%\bibliography{bib}

\end{document}